\documentclass[fleqn,usenatbib]{mnras}

\usepackage{newtxtext,newtxmath}

\usepackage[T1]{fontenc}

\DeclareRobustCommand{\VAN}[3]{#2}
\let\VANthebibliography\thebibliography
\def\thebibliography{\DeclareRobustCommand{\VAN}[3]{##3}\VANthebibliography}

\usepackage{graphicx}	
\usepackage{amsmath}	
\usepackage{tabularray}
\usepackage{longtable}
\usepackage{ltablex}
\usepackage{booktabs}
\usepackage{natbib} 
\usepackage{float}
\keepXColumns

\title[Quasar--Companion System Without AGN Outflow]{A Quasar--Companion System Without AGN Outflow at $z \sim 6$: The Case of PSO J083+11}

\author[M. A. Husain et al.]{
	Muhammad Akmal Husain,$^{1,2}$\thanks{E-mail: 20924010@mahasiswa.itb.ac.id}
	Irham Taufik Andika,$^{3,4,5}$\thanks{E-mail: irham.andika@lmu.de}
	and Mochamad Ikbal Arifyanto$^{2,6,7}$\thanks{E-mail: mochamad.ikbal@itb.ac.id}
	\\
	$^{1}$Computational Science Graduate Program, Faculty of Mathematics and Natural Sciences, Bandung Institute of Technology, Bandung 40132, Indonesia\\
	$^{2}$Department of Astronomy, Faculty of Mathematics and Natural Sciences, Bandung Institute of Technology, Bandung 40132, Indonesia\\
	$^{3}$University Observatory, LMU Faculty of Physics, Scheinerstrasse 1, 81679 Munich, Germany\\
	$^{4}$Technical University of Munich, TUM School of Natural Sciences, Department of Physics, James-Franck-Str. 1, 85748 Garching, Germany\\
	$^{5}$Max-Planck-Institut f\"{u}r Astrophysik, Karl-Schwarzschild-Str. 1, 85748 Garching, Germany\\
	$^{6}$Astronomy Research Group, Faculty of Mathematics and Natural Sciences, Bandung Institute of Technology, Bandung 40132, Indonesia\\
	$^{7}$Bosscha Observatory, Bandung Institute of Technology, Jl. Peneropongan Bintang No. 1, 40391, Indonesia
}

\date{Accepted 2026 April 21. Received 2026 March 13; in original form 2025 November 11}

\pubyear{2025}

\begin{document}
	\label{firstpage}
	\pagerange{\pageref{firstpage}--\pageref{lastpage}}
	\maketitle
	
	\begin{abstract}
		PSO J083.8371+11.8482, a quasar at $z = 6.34$ with a nearby companion galaxy, provides an opportunity to study the impact of active galactic nucleus (AGN) activity on the surrounding environment during the epoch of reionization. We analyze ALMA observations of the [C\,\textsc{ii}] 158~$\mu$m emission line and the far-infrared (FIR) continuum, which trace cold interstellar gas and dust-reprocessed radiation from star formation and AGN heating. The quasar host shows star formation rates (SFRs) of $544$--$3764~\mathrm{M_{\odot}~yr^{-1}}$ from [C\,\textsc{ii}] and $1861$--$2932~\mathrm{M_{\odot}~yr^{-1}}$ from FIR emission, while the companion galaxy exhibits lower SFRs of $21$--$145$ and $76$--$211~\mathrm{M_{\odot}~yr^{-1}}$ from the same diagnostics. Both galaxies follow typical $L_{\mathrm{[C\,II]}}/L_{\mathrm{FIR}}$ ratios observed in star-forming galaxies and show no evidence for a [C\,\textsc{ii}] deficit, indicating that stellar heating dominates the interstellar medium energetics. The [C\,\textsc{ii}] moment maps reveal compact emission with centrally peaked intensity and ordered rotational kinematics in both systems. Velocity dispersions remain well below values associated with powerful AGN-driven outflows, and no significant morphological asymmetries or disturbed velocity fields indicative of AGN feedback or major mergers are detected, although marginal kinematic substructure in the quasar's high-velocity channels warrants further investigation. Although the companion lies at a projected distance of $18.248 \pm 0.277$~kpc within the quasar proximity zone, neither morphological nor kinematic signatures indicate AGN-driven outflows affecting the circumgalactic medium. We therefore interpret this system as being observed in a pre-outflow accretion phase, where rapid supermassive black hole growth precedes the development of large-scale AGN feedback.
	\end{abstract}
	
	
	
	
	\section{Introduction}
	
	Quasi-stellar radio sources (quasars) are non-transient objects that belong to the unified model of active galactic nuclei (AGN). supermassive black holes (SMBHs) that accrete matter from their host galaxies \citep{Netzer2015, Padovani2017, Inayoshi2019} power these luminous objects. This accretion process releases an enormous amount of energy across nearly the entire electromagnetic spectrum, enabling quasars to appear as bright beacons along cosmological distances \citep{Fan2006, Banados2018}. This allows us to investigate the early universe, the formation of large-scale structures, and the cosmic reionization ($z > 6$) process more comprehensively \citep{BrommYoshida2011, Mortlock2011, Wang2021}.
	
	During the reionization era, quasars contributed to the ionization of the intergalactic medium (IGM), resulting in the formation of significant ionized bubbles known as proximity zones \citep{Haiman1999, Madau2000, Mesinger2016, Eilers2017, Davies2019}. The characteristics of these zones, including their size and ionization profiles, provide insight into the luminosity, lifespan, and radiation feedback of primordial quasars.
	
	AGNs, including quasars, play an important role in galaxy evolution through a process called AGN feedback. Observational and numerical simulations show that AGN feedback modulates gas flows and star formation by transferring energy and momentum to the interstellar medium (ISM) and the surrounding circumgalactic environment \citep{Hopkins2010, Fabian2012, Harrison2017}. AGN feedback is generally divided into two categories. Quasar-mode feedback is characterized primarily by strong radiation from the central AGN, which can drive radiatively-powered winds \citep{Fabian2012, Harrison2017}. In contrast, radio-mode feedback involves powerful jets and large-scale outflows that mechanically heat and redistribute gas in the galaxy's environment \citep{Cicone2014, Zubovas2014, Costa2015}. 
	
	The impact of AGN feedback on star formation can be either positive or negative, depending on environmental factors and the specific physical conditions in the host galaxy. Positive feedback can enhance star formation by compressing cold gas through shocks and pressure from AGN-driven outflows \citep{Zinn2013, Zubovas2013}. Conversely, negative feedback reduces star formation by heating the ISM, expelling gas reservoirs, or preventing cooling flows in massive systems \citep{Schawinski2006, Dubois2013}. The balance between these competing processes is influenced by factors such as gas density, AGN luminosity, host galaxy mass, and the system's evolutionary state \citep{Bicknell2010, Shabala2010, Ishibashi2012, Morganti2017}.
	
	Continuum diagnostics and fine-structure lines in the far-infrared (FIR) are one of the observational methods for identifying signs of AGN feedback. The [C\,\textsc{ii}] 158\,$\mu$m line, formed by singly ionized carbon (C$^+$), is the primary cooler of the diffuse ISM and photodissociation regions (PDRs). This line is frequently used to identify gas kinematics, star formation rates (SFR), and feedback effects in high-redshift galaxies \citep{Stacey1991, DeLooze2014, HerreraCamus2015, Lagache2018, Zanella2018}. The [C\,\textsc{ii}] line is often analyzed alongside the continuous FIR emission from dust, which reflects the integrated heating from stars and AGNs within the galaxy \citep{CarilliWalter2013, Casey2014, Gruppioni2020}. Thus, the ratio of these two quantities ($L_{\mathrm{[C\,II]}}/L_{\mathrm{FIR}}$) serves as a strong diagnostic for heating processes within the host galaxy. In typical star-forming galaxies, this ratio remains relatively constant at $\sim 10^{-3}$ \citep{Malhotra2001, Stacey2010}, whereas in compact starburst galaxies or AGN-dominated systems, it often decreases significantly relative to this normal value. This phenomenon is known as the ``[C\,\textsc{ii}] deficit'' \citep{Luhman2003, Gullberg2015, Munoz2016}. This deficit can be produced by several mechanisms, including high ionization parameters (i.e., the ratio of ionizing photon density to gas density, which can be elevated by intense AGN radiation), dust-bound H\,\textsc{ii} regions, or additional ionization of C$^+$ to C$^{2+}$ by hard AGN radiation \citep{Narayanan2017, Ebagezio2024}.
	
	The latest Atacama Large Millimeter/submillimeter Array (ALMA) observations of luminous quasars and their environments support the influence of AGNs on [C\,\textsc{ii}] and FIR emissions. Radiative heating from AGNs increases FIR dust luminosity while reducing [C\,\textsc{ii}] emissions by neutralizing C$^+$ or lowering photoelectric heating efficiency. As a result, these effects may obscure the characteristic signs of star formation and complicate the interpretation of SFR from FIR diagnostics \citep{Croxall2012, Sargsyan2012, Decarli2018, Carniani2020}. This can lead to an artificial [C\,\textsc{ii}] deficit even during periods of active star formation, highlighting the importance of spatially resolved observations to separate AGN-induced heating from stellar processes.
	
	Companion galaxies, found within a few tens of kiloparsecs of high-redshift quasars, serve as natural laboratories for testing the extent and strength of AGN feedback beyond the main quasar host. These companion galaxies are likely susceptible to heating or gas compression caused by the quasar’s radiation field and outflows, depending on their position and the orientation of AGN-driven phenomena \citep{Yue2019, Zana2022}. 
	
	Contrary to earlier assumptions, recent observations reveal that high-redshift quasars often have companion galaxies. ALMA observations have uncovered a high fraction of companion galaxies within projected separations of $< 100$ kpc from $z \sim 6$ quasars, supporting the idea that the first generation of high-mass, luminous SMBHs grew in over-dense environments \citep{Trakhtenbrot2019}. Theoretical and numerical studies using cosmological hydrodynamical simulations have demonstrated the existence not only of single SMBHs but also of multiple SMBHs (binary or triplet systems) at the centers of  $z \sim 6$ AGN \citep{Barai2018}. More recently, James Webb Space Telescope (JWST) observations have enabled the detection of companion galaxies around high-z quasars with unprecedented detail, revealing complex merging systems and providing new insights into the rapid growth of early SMBHs and their host galaxies \citep{Decarli2024}. 
	
	When observed, companion galaxies often exhibit disturbed gas morphology, compact FIR structures, or increased SFR, which may be caused by gravitational interactions or AGN impacts \citep{Neeleman2019, BPVenemans2020, Novak2020}.
	
	We use ALMA observations in the FIR continuum and the [C\,\textsc{ii}] 158\,$\mu$m line to study signs of AGN feedback in the PSO J083.8371+11.8482 system at $z \sim 6.34$. We aim to determine whether the quasar's energy output affects not only its host galaxy but also the gas conditions and star formation in its closest companion. We analyze moment maps, [C\,\textsc{ii}] line profiles, and the $L_{\mathrm{[C\,II]}}/L_{\mathrm{FIR}}$ ratio to detect potential signals from AGN-driven outflows, radiative heating, or dynamical disturbances. This study builds on the latest ALMA campaign, which discovered similar feedback mechanisms in high-redshift quasar systems and contributes to a deeper understanding of the role of AGNs in shaping galaxy evolution during the reionization era \citep{Decarli2018, BPVenemans2020, Novak2020}.
	
	The structure of this paper is as follows. In Section~\ref{sec:Data}, we describe the data acquisition and processing. Section~\ref{sec:Methods} presents our FIR spectral measurements and analysis methods. The results, including our examination of AGN feedback, are detailed in Section~\ref{sec:Results} and further discussed in Section~\ref{sec:Discussion}. Finally, we summarize our main conclusions in Section~\ref{sec:Conclusion}.
	
	\section{Data}
	\label{sec:Data}
	
	The primary object of this study is the quasar PSO J083.8371+11.8482, first discovered through the optical survey Pan-STARRS1 \citep{Andika2020}. Based on these observations, a companion galaxy was identified southwest of the quasar at coordinates $\alpha = 83.83727^\circ$ and $\delta = +11.8474^\circ$. This companion galaxy is suspected to be gravitationally linked to the quasar PSO J083+11, making it a crucial focus in analyzing gravitational interactions and the effects of AGN feedback.
	
	Observations for this study were conducted on October 9, 2019, using the ALMA in the C43-4 configuration, with a total on-source integration time of 3145 seconds. The spectral setup was optimized to target the redshifted [C\,\textsc{ii}] 158\,$\mu$m line, centering around $\sim$258\,GHz, which corresponds to its rest-frame frequency of 1900.5369\,GHz at $z = 6.34$. The dataset includes both [C\,\textsc{ii}] emission line spectra and FIR continuum measurements, which are critical tracers of interstellar gas conditions and star formation in galaxies. Data reduction and image reconstruction were performed using the \texttt{tclean} function within the Common Astronomy Software Applications (\texttt{CASA}) package, applying gridding, Briggs weighting (to balance sensitivity and resolution), and deconvolution to enhance image fidelity \citep{McMullin2007, The_CASA_Team_2022}.
	
	The cleaning process was limited to a 5\,arcsec circular region centered on the quasar, with a cleaning threshold of 0.44\,mJy, established by the root mean square (RMS) oscillations in the backdrop of the filthy image. The RMS noise level, defined as the square root of the mean of the squared pixel intensities in signal-free regions, is a standard metric for quantifying thermal and instrumental noise in radio interferometric imaging. It is commonly used to assess image sensitivity and define signal detection thresholds \citep{rau2011msmf}. The final data cube, constructed with a spectral resolution of 30\,MHz per channel, achieved an RMS noise level of roughly 0.24\,mJy\,beam$^{-1}$. This indicates that features over $\sim$0.72\,mJy\,beam$^{-1}$ ($3\sigma$) are statistically significant. Spectral extraction was performed using elliptical apertures of 3.26\arcsec\ $\times$ 2.73\arcsec\ for the quasar and 0.96 \arcsec\ $\times$ 1.21 \arcsec\ for the companion galaxy.
	
	\section{Methods}
	\label{sec:Methods}
	
	All analyses in this work adopt a flat $\Lambda$CDM cosmology with $H_0 = 70~\mathrm{km~s^{-1}~Mpc^{-1}}$, $\Omega_\Lambda = 0.7$, $\Omega_\mathrm{m} = 0.3$, and $\Omega_\mathrm{b} = 0.05$ \citep{Planck2020}. The physical parameters of our sources are determined via spectral analysis of ALMA data. As [C\,\textsc{ii}] 158\,$\mu$m and the FIR continuum are sensitive diagnostics of ISM conditions and star formation in galaxies, we use tailored methods for robust continuum subtraction and line profile fitting, following best-practice procedures recommended for extragalactic (sub)mm data \citep{Leroy2015}.
	
	\subsection{[C\,\textsc{ii}] 158\,$\mu$m Emission Line Extraction and Fitting}
	\label{sec:cii_fitting}
	
	The fine structure line [C\,\textsc{ii}] 158\,$\mu$m provides a crucial diagnostic tool for investigating the characteristics of the ISM and star formation processes throughout cosmic history, from local galaxy systems to high redshift objects \citep{Stacey2010, CarilliWalter2013, DeLooze2014, HerreraCamus2015, Carniani2018}. We employ a robust analytical approach to extract this line reliably from ALMA data.
	
	Continuum estimation and subtraction using an iterative sigma-clipping approach \citep{Feigelson2012, astropy:2013, astropy:2018, astropy:2022}, which systematically discards spectral channels exceeding a $3\sigma$ until convergence is achieved. The continuum level is determined as the median of the retained channels, with uncertainty calculated using standard statistical methods. This baseline is then subtracted to isolate the [C\,\textsc{ii}] emission line.
	
	The extracted lines are modeled using a single Gaussian profile using the Levenberg--Marquardt optimization routine \citep{astropy:2013, astropy:2018, astropy:2022, Virtanen2020}, with adjustments for amplitude, center velocity, and velocity width. Parameter uncertainties are obtained from the covariance matrix, with residual-based estimation used for low S/N cases. Line flux is determined through trapezoidal numerical integration on the continuum-corrected spectrum, accounting for possible asymmetry or double components. Flux uncertainties include both spectral noise and integration bandwidth effects using the mean square error (MSE) propagation method.
	
	From the Gaussian fit, we extract the systemic velocity and velocity dispersion of the source, which are useful for interpreting the galaxy's kinematics. Following \citet{Andika2020}, the integrated [C\,\textsc{ii}] flux is converted into luminosity using the standard relation \citep{CarilliWalter2013}:
	\begin{equation}
		\frac{L_{\rm [C\,II]}}{L_\odot} = 1.04 \times 10^{-3} 
		\frac{S_{\text{line}} \Delta v}{\mathrm{Jy~km~s}^{-1}}
		\left( \frac{D_L}{\mathrm{Mpc}} \right)^2 \nu_{\mathrm{obs}}~(\mathrm{GHz}),
	\end{equation}
	where $D_L$ is the luminosity distance and $\nu_{\mathrm{obs}}$ is the observed frequency of the line.
	
	Finally, we estimate the SFR using the empirical [C\,\textsc{ii}]--SFR relation from \citet{DeLooze2014}:
	\begin{equation}
		\log \mathrm{SFR}_{\rm [C\,II]} = -6.99 + 1.01 \log L_{\rm [C\,II]},
	\end{equation}
	which has been shown to hold across a wide range of galaxy populations, including main-sequence star-forming galaxies, luminous infrared galaxies, and high-redshift sources \citep{HerreraCamus2015, Carniani2018, Schaerer2020}.
	
	\subsection{FIR Emission Fitting}
	\label{sec:fir_fitting}
	
	We modeled the dust emission, which is represented by the FIR continuum, using a modified blackbody function. Only the dust mass, $M_{\text{dust}}$, was allowed to vary during fitting. Following \citet{Beelen2006} and \citet{Novak2019}, the FIR flux density is expressed as:
	
	\begin{equation}
		S_{\nu_{\text{obs}}} = f_{\text{cmb}} (1 + z) D_L^{-2} \kappa_{\nu_\text{rest}} M_{\text{dust}} B_{\nu_\text{rest}} (T_{\text{dust}, z}),
		\label{eqBeelenNovak}
	\end{equation}
	
	where $B_{\nu_\text{rest}}(T)$ is the Planck blackbody function, $\kappa_{\nu_\text{rest}}$ is the dust opacity coefficient, and $f_{\text{cmb}}$ corrects for contamination from the Cosmic Microwave Background (CMB) at high redshift. This correction becomes especially significant at $z > 6$, where the CMB temperature approaches that of cold dust. The CMB correction and dust temperature evolution are calculated as follows \citep{daCunha2013}:
	
	\begin{equation}
		f_{\text{cmb}} = 1 - \frac{B_{\nu \text{rest}} (T_{\text{cmb}, z})}{B_{\nu \text{rest}} (T_{\text{dust}, z})},
		\label{daCunha1}
	\end{equation}
	\begin{equation}
		T_{\text{dust}, z} = \left( T_{\text{dust}}^{\beta + 4} + T_{\text{cmb}, z=0}^{\beta + 4} \left[(1 + z)^{\beta + 4} - 1\right] \right)^{\frac{1}{\beta + 4}},
		\label{daCunha2}
	\end{equation}
	\begin{equation}
		T_{\text{cmb}, z} = 2.73 \times (1 + z).
		\label{daCunha3}
	\end{equation}
	
	We adopted a fiducial dust temperature of $T_{\text{dust}} = 47$\,K and an emissivity index $\beta = 1.6$, consistent with high-redshift quasar studies \citep{Dunne2003,Beelen2006}. The dust opacity $\kappa_{\nu_{\text{rest}}}$ is frequency-dependent and follows:
	
	\begin{equation}
		\kappa_{\nu_\text{rest}} = \kappa_{\nu_0} \left( \frac{\nu_{\text{rest}}}{\nu_0} \right)^{\beta} = 2.64 \left( \frac{\nu_{\text{rest}}}{c/125\,\mu\mathrm{m}} \right)^{\beta} \, \text{m}^2 \text{kg}^{-1},
		\label{BeelenNovak2}
	\end{equation}
	\begin{equation}
		\nu_\text{rest} = \nu_\text{obs} (1 + z).
		\label{BeelenNovak3}
	\end{equation}
	
	The Planck function $B_{\nu_{\text{rest}}}$ in Eq.~\ref{eqBeelenNovak} is defined as:
	
	\begin{equation}
		B_{\nu_\text{rest}}({\nu_\text{rest}}, T) = \frac{2h{\nu_\text{rest}}^3}{c^2} \cdot \frac{1}{e^{\frac{h{\nu_\text{rest}}}{k_B T}} - 1},
		\label{Planck}
	\end{equation}
	
	where $h = 6.626 \times 10^{-34}~\text{J\,s}$ is Planck's constant, and $k_B = 1.381 \times 10^{-23}~\text{J\,K}^{-1}$ is Boltzmann's constant.
	
	After fitting the FIR spectral energy distribution using the \texttt{lmfit} package \citep{Newville2014}, we numerically integrated the best-fit curve over the rest-frame wavelength interval 42.5--122.5\,$\mu$m to obtain the total FIR luminosity, $L_{\text{FIR}}$. The obscured SFR was then calculated using the FIR-based SFR conversion from \citet{Kennicutt1998}:
	
	\begin{equation}
		\text{SFR} (\mathrm{M}_\odot\,\mathrm{yr}^{-1}) = 4.5 \times 10^{-44} \times L_{\text{FIR}}\,(\mathrm{erg\,s}^{-1}).
		\label{eqSFRFIR}
	\end{equation}
	
	Although this relation is commonly used for star-forming galaxies, in systems hosting powerful AGN, the FIR continuum may contain a non-negligible contribution from AGN-heated dust. Therefore, we interpret the SFR$_\text{FIR}$ values with caution and compare them against [C\,\textsc{ii}]-based estimates to account for potential contamination and heating bias \citep{Sargsyan2012, Carniani2020}.
	
	\subsection{Evaluation of AGN Feedback}
	\label{sec:agn_feedback}
	
	To investigate potential AGN feedback signatures in PSO J083+11 and its companion galaxy, we employ a multi-panel comparative diagnostic approach utilizing the relationship between [C\,\textsc{ii}] 158~$\mu$m and FIR continuum emission, following the framework established by \citet{Decarli2018} and \citet{BPVenemans2020} for high-redshift quasar systems.
	
	We construct three complementary diagnostics presented in Figures~\ref{fig:cii_deficit} and \ref{fig:Hasil_Akhir}. The first diagnostic (Figure~\ref{fig:cii_deficit}, left panel) plots the luminosity ratio $L_{\text{[C\,II]}}/L_{\text{FIR}}$ against total FIR luminosity, comparing our targets against diverse galaxy populations spanning $z < 0.2$ to $z \sim 7$ \citep{Stacey2010, Gullberg2015, Hughes2017, BPVenemans2020}. The second diagnostic (Figure~\ref{fig:cii_deficit}, right panel) plots $L_{\text{[C\,II]}}/L_{\text{FIR}}$ against FIR surface brightness $\Sigma_{\text{FIR}}$, providing a spatially-sensitive metric connecting to the compactness and intensity of the radiation field heating the dust \citep{Lutz2016, DiazSantos2017}. We calculate $\Sigma_{\text{FIR}}$ values using the empirical size-luminosity relation from \citet{Lutz2016} based on spatially resolved observations of local star-forming galaxies and quasars \citep{Decarli2018}. Two theoretical models are overlaid for comparison. The \citet{Lutz2016} model (solid line) represents systems where stellar processes dominate dust heating, while the \citet{DiazSantos2017} model (dashed line) characterizes compact starburst environments with intense radiation fields \citep{DiazSantos2017, Decarli2018}. The third diagnostic (Figure~\ref{fig:Hasil_Akhir}) displays the direct correlation between [C\,\textsc{ii}] and FIR luminosities for our comparison sample, showing the best-fit regression and 1$\sigma$ dispersion \citep{Stacey2010, DeLooze2014, Carniani2018, Schaerer2020}.
	
	Our comparison sample (Table~\ref{tab:galaxy_sample_extended}) comprises 105 galaxies spanning local to high-redshift systems across multiple galaxy populations and evolutionary stages. For $\Sigma_{\text{FIR}}$ measurements, we adopt values estimated using the empirical size-luminosity relation from \citet{Lutz2016} for the majority of sources, and direct measurements from \citet{Decarli2018} for $z \sim 6$ quasars where spatially resolved observations are available. The uncertainty in $L_{\text{[C\,II]}}$ measurements includes contributions from both observational noise and systematic uncertainties. Following \citet{DeLooze2014}, we adopt an intrinsic scatter of approximately 0.42~dex in the [C\,\textsc{ii}] to SFR relation. Uncertainties in $L_{\text{FIR}}$ are derived from the dust mass fitting procedure (Section~\ref{sec:fir_fitting}), while uncertainties in $\Sigma_{\text{FIR}}$ incorporate both luminosity uncertainties and the scatter in the size-luminosity relation from \citet{Lutz2016}.
	
	\section{Results}
	\label{sec:Results}
	
	\subsection{Fitting Result}
	
	\begin{figure*}
		\centering
		\includegraphics[width=0.86\columnwidth]{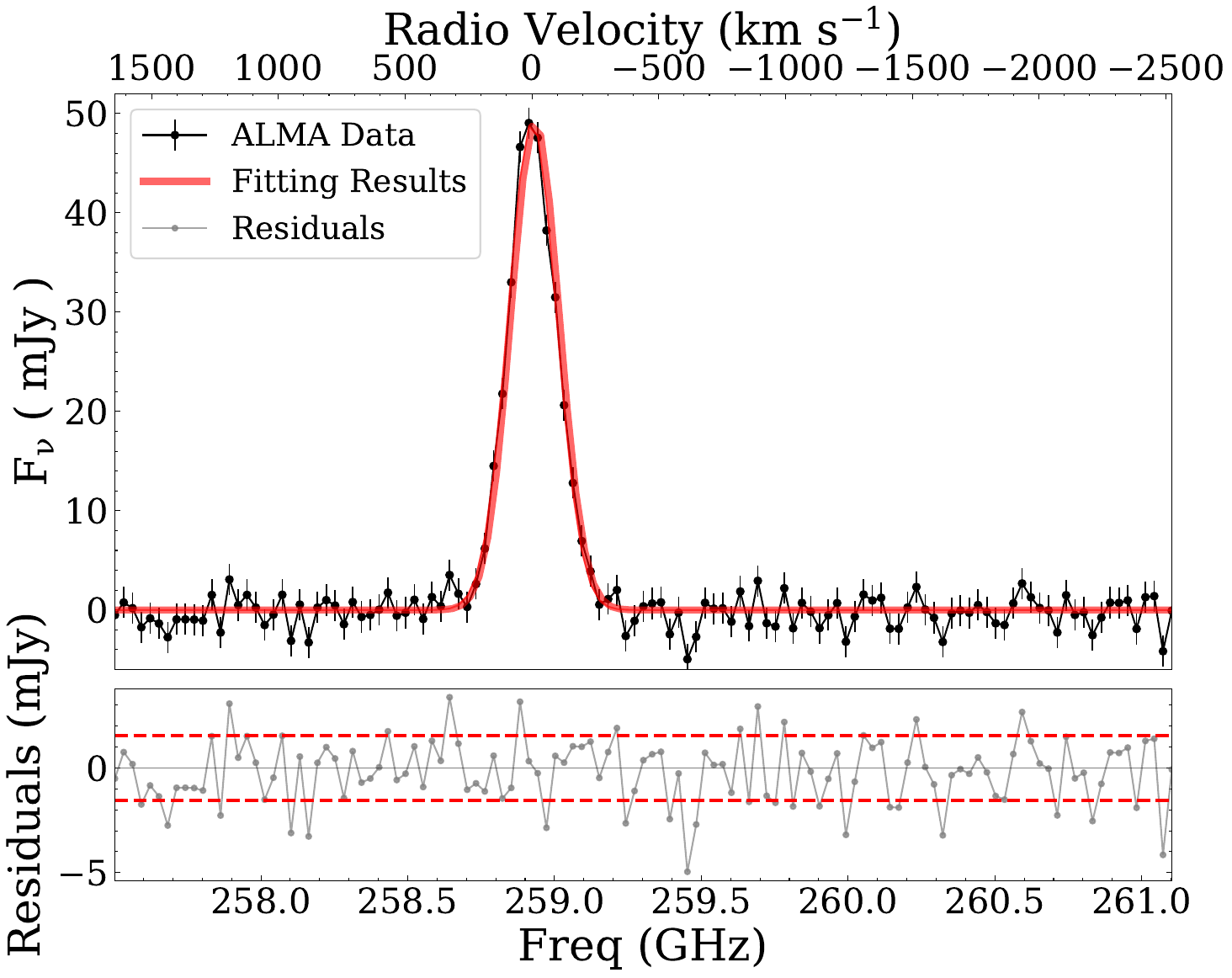}
		\includegraphics[width=\columnwidth]{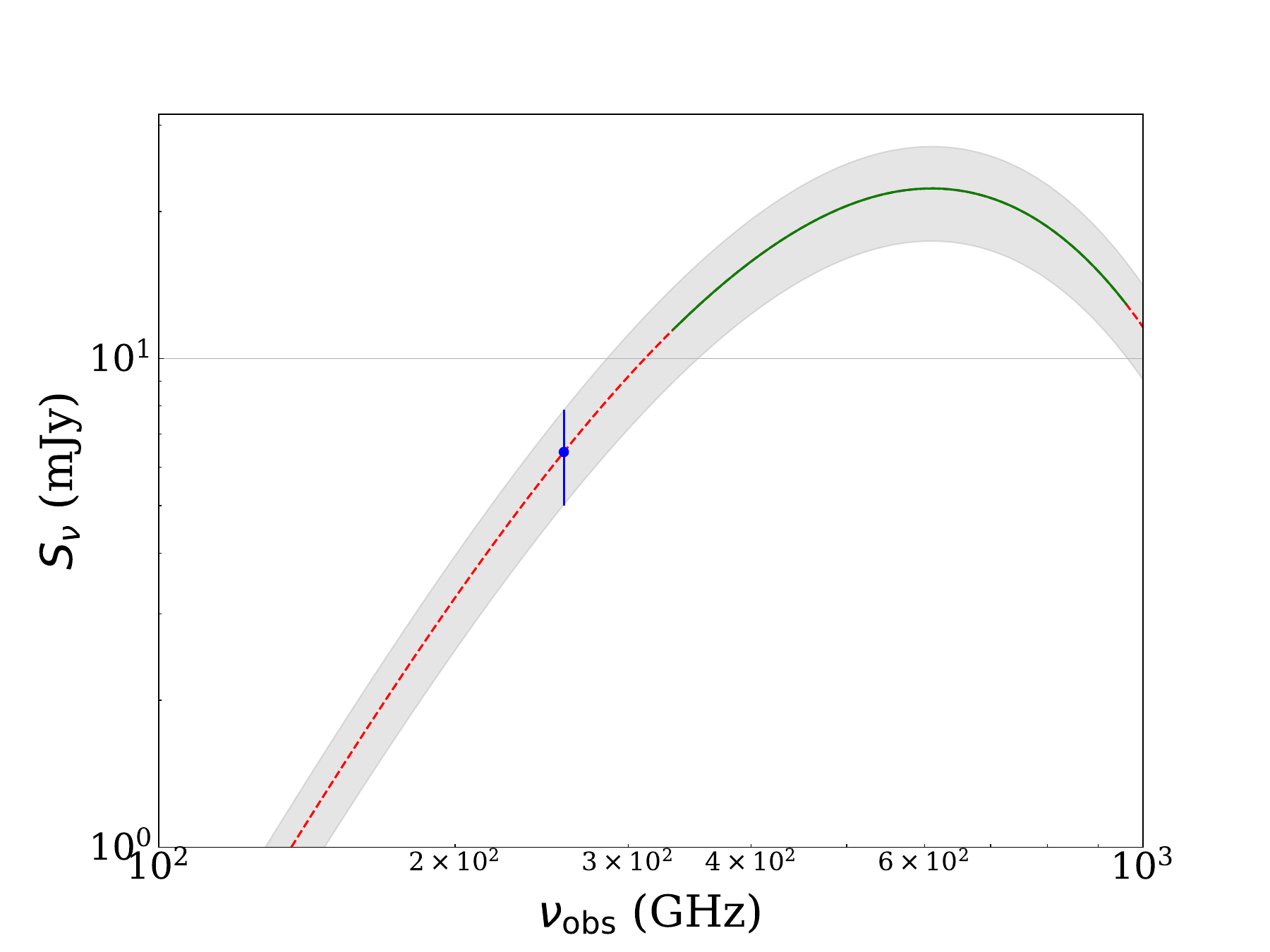}
		\includegraphics[width=0.86\columnwidth]{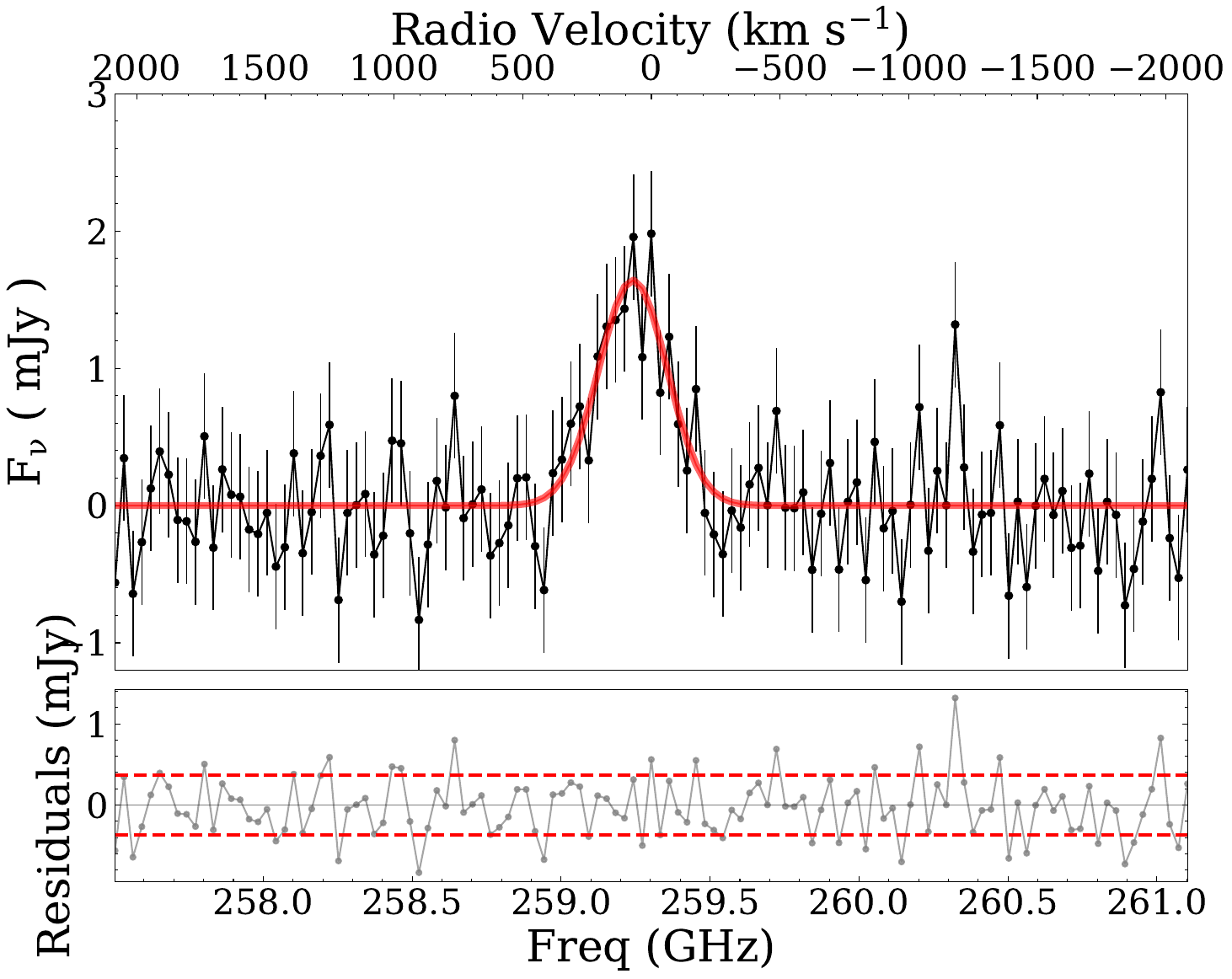}
		\includegraphics[width=\columnwidth]{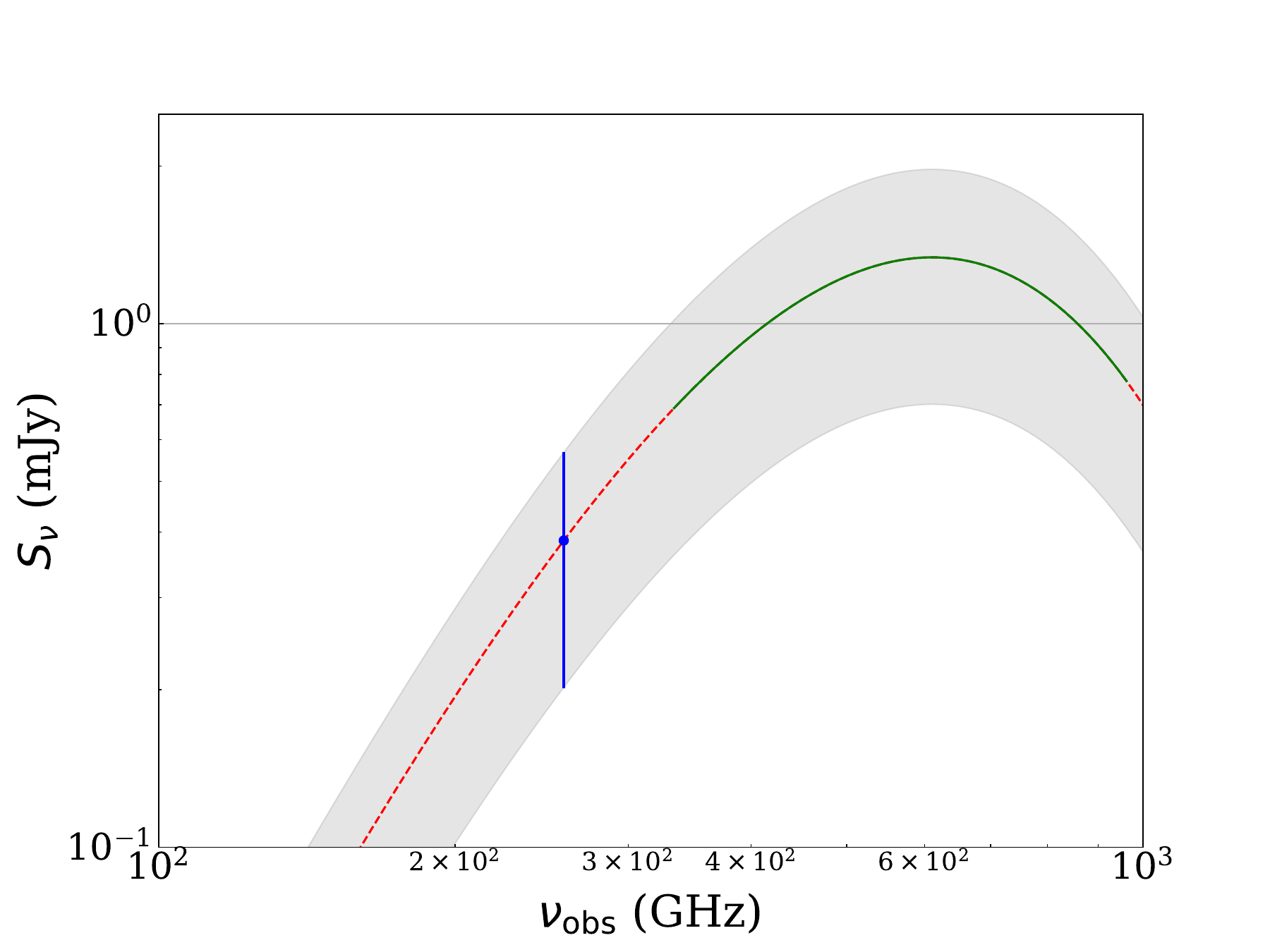}
		\caption{Spectral fitting results for PSO J083+11 and its companion galaxy. (Upper left panel) [C\,\textsc{ii}]~158~$\mu$m emission line spectrum of the quasar, with observed data (black points with error bars) and best-fit Gaussian profile (red curve). (Lower left panel) [C\,\textsc{ii}] spectrum of the companion galaxy with the same representation. (Upper right panel) FIR spectral energy distribution of the quasar, showing observational data (blue points), the FIR wavelength integration range (green line), and Total-Infrared contribution (red dashed line). (Lower right panel) FIR spectral energy distribution of the companion galaxy with identical notation.}
		\label{fig:quasar_companion_fitting}
	\end{figure*}
	
	\begin{table}
		\caption{Summary of FIR and [C\,II] Fitting Results for the Quasar and Companion Galaxy}
		\label{tab:tablequasarparam}
		\begin{tabular}{lrc}
			\hline
			Parameter & Value & Unit \\
			\hline
			\multicolumn{3}{c}{\textbf{Quasar}} \\
			$z_{\mathrm{[C\,II]}}$ & $6.3401 \pm 0.0004$ & --- \\
			$D_{L}(z_{\mathrm{[C\,II]}})$ & $61547.052 \pm 3.546$ & Mpc \\
			$A$ & $49.138 \pm 1.565$ & mJy \\
			FWHM ([C\,II]) & $224.467 \pm 22.447$ & km\,s$^{-1}$ \\
			Reduced-$\chi^{2}$ ([C\,II]) & 1.014 & --- \\
			$\nu_{\mathrm{obs}}$ & $258.924 \pm 0.001$ & GHz \\
			Flux ([C\,II]) & $10.878 \pm 0.641$ & Jy\,km\,s$^{-1}$ \\
			$S_{258.9\,\mathrm{GHz}}$ & $6.44 \pm 1.430$ & mJy \\
			$S_{\mathrm{total}}$ (FIR) & $(1.167 \pm 0.260) \times 10^{10}$ & Jy \\
			$L_{\mathrm{[C\,II]}}$ & $(1.110 \pm 0.065) \times 10^{10}$ & $L_\odot$ \\
			$L_{\mathrm{FIR}}$ & $(1.382 \pm 0.307) \times 10^{13}$ & $L_\odot$ \\
			log($L_{\mathrm{[C\,II]}}/L_{\mathrm{FIR}}$) & $-3.095 \pm 0.099$ & --- \\
			log($\Sigma_{\mathrm{FIR}}$) & $13.418 \pm 0.270$ & $L_{\odot}$~kpc$^{-2}$ \\
			SFR$_{\mathrm{[C\,II]}}$ & $544.062$~--~$3763.988$ & $M_\odot$\,yr$^{-1}$ \\
			SFR$_{\mathrm{FIR}}$ & $1860.734$~--~$2932.462$ & $M_\odot$\,yr$^{-1}$ \\
			$M_{\mathrm{dust}}$ & $(5.532 \pm 1.229) \times 10^{8}$ & $M_\odot$ \\
			\hline
			\multicolumn{3}{c}{\textbf{Companion Galaxy}} \\
			$z_{\mathrm{[C\,II]}}$ & $6.3309 \pm 0.0004$ & --- \\
			$D_{L}(z_{\mathrm{[C\,II]}})$ & $61443.444 \pm 3.545$ & Mpc \\
			$A$ & $1.644 \pm 0.369$ & mJy \\
			FWHM ([C\,II]) & $313.126 \pm 31.313$ & km\,s$^{-1}$ \\
			Reduced-$\chi^{2}$ ([C\,II]) & 0.671 & --- \\
			$\nu_{\mathrm{obs}}$ & $259.240 \pm 0.005$ & GHz \\
			Flux ([C\,II]) & $0.434 \pm 0.151$ & Jy\,km\,s$^{-1}$ \\
			$S_{259.2\,\mathrm{GHz}}$ & $0.385 \pm 0.184$ & mJy \\
			$S_{\mathrm{total}}$ (FIR) & $(7.016 \pm 3.320) \times 10^{8}$ & Jy \\
			$L_{\mathrm{[C\,II]}}$ & $(4.442 \pm 1.542) \times 10^{8}$ & $L_\odot$ \\
			$L_{\mathrm{FIR}}$ & $(8.283 \pm 3.920) \times 10^{11}$ & $L_\odot$ \\
			log($L_{\mathrm{[C\,II]}}/L_{\mathrm{FIR}}$) & $-3.272 \pm 0.255$ & --- \\
			log($\Sigma_{\mathrm{FIR}}$) & $11.698 \pm 0.328$ & $L_{\odot}$~kpc$^{-2}$ \\
			SFR$_{\mathrm{[C\,II]}}$ & $20.992$~--~$145.222$ & $M_\odot$\,yr$^{-1}$ \\
			SFR$_{\mathrm{FIR}}$ & $75.518$~--~$211.190$ & $M_\odot$\,yr$^{-1}$ \\
			$M_{\mathrm{dust}}$ & $(3.315 \pm 1.569) \times 10^{7}$ & $M_\odot$ \\
			\hline
		\end{tabular}
	\end{table}
	
	The spectral fitting results for PSO J083+11 and its companion galaxy are summarized in Table~\ref{tab:tablequasarparam} and illustrated in Figure~\ref{fig:quasar_companion_fitting}. The [C\,\textsc{ii}] emission line spectrum of the quasar (Upper left panel) exhibits a well-defined single-peaked profile with a reduced $\chi^2$ value near unity, indicating an excellent fit to the observed data with minimal residual structure. The symmetric, single-component line profile shows no evidence of asymmetric wings, double-peaked structure, or broad velocity components that would indicate large-scale outflows or complex kinematics \citep{Walter2004, Carniani2013, Carniani2018}. The companion galaxy's [C\,\textsc{ii}] emission (Lower left panel) is similarly well-described by a single Gaussian component without significant deviations from a smooth symmetric profile.
	
	Both systems exhibit regular, symmetric [C\,\textsc{ii}] line profiles that are well-fitted by single Gaussian functions. The derived line widths fall within the range observed for $z > 6$ galaxies dominated by rotational support rather than velocity dispersion \citep{Neeleman2019, Pensabene2020, Schouws2022}. The absence of broad wings, secondary components, or significant residuals in the spectral fits is characteristic of morphologically undisturbed systems \citep{Carniani2018, BPVenemans2020}.
	
	The FIR spectral energy distribution for both systems (right panel) fits well, so that the calculated dust mass and FIR luminosity are consistent with values reported for similar high-redshift galaxies. The quasar dust mass is comparable to values found in other quasar hosts with $z \sim 6$ \citep{Beelen2006, BPVenemans2020, Mazzucchelli2025}, while the companion dust mass is within the range observed in satellite galaxies around high-redshift quasars \citep{Decarli2018, BPVenemans2020}. 
	
	The SFR estimates derived from both [C\,\textsc{ii}] and FIR tracers place PSO J083+11 in the regime of intensely star-forming systems at $z \sim 6$ \citep{Carniani2018, Decarli2018, BPVenemans2020}, while the companion exhibits star formation activity typical of satellite galaxies in overdense environments \citep{Neeleman2019, BPVenemans2020}. The range in SFR estimates reflects the intrinsic scatter of 0.42~dex in the [C\,\textsc{ii}] to SFR calibration relation from \citet{DeLooze2014}.
	
	\subsection{Diagnostic Plot}
	
	\begin{figure*}
		\centering
		\includegraphics[width=2\columnwidth]{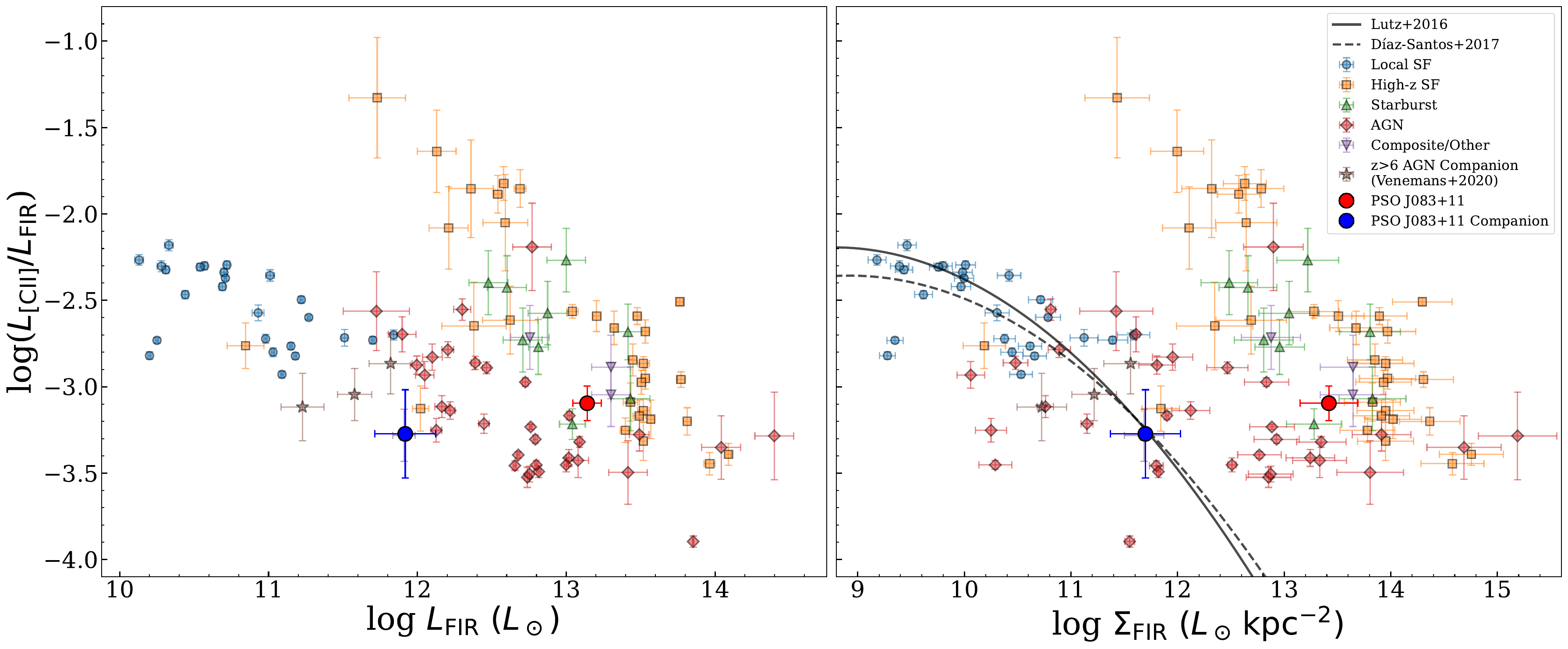}
		\caption{Dual diagnostic for [C\,\textsc{ii}] deficit evaluation. (Left panel) $L_{\text{[C\,II]}}/L_{\text{FIR}}$ versus FIR luminosity. (Right panel) $L_{\text{[C\,II]}}/L_{\text{FIR}}$ versus FIR surface brightness $\Sigma_{\text{FIR}}$, with theoretical models from \citet{Lutz2016} (solid line, derived from local star-forming galaxies and local quasars) and \citet{DiazSantos2017} (dashed line, derived from local ULIRGs and PDR simulations). The comparison sample is detailed in Table~\ref{tab:galaxy_sample_extended}.}
		\label{fig:cii_deficit}
	\end{figure*}
	
	\begin{figure}
		\centering
		\includegraphics[width=\columnwidth]{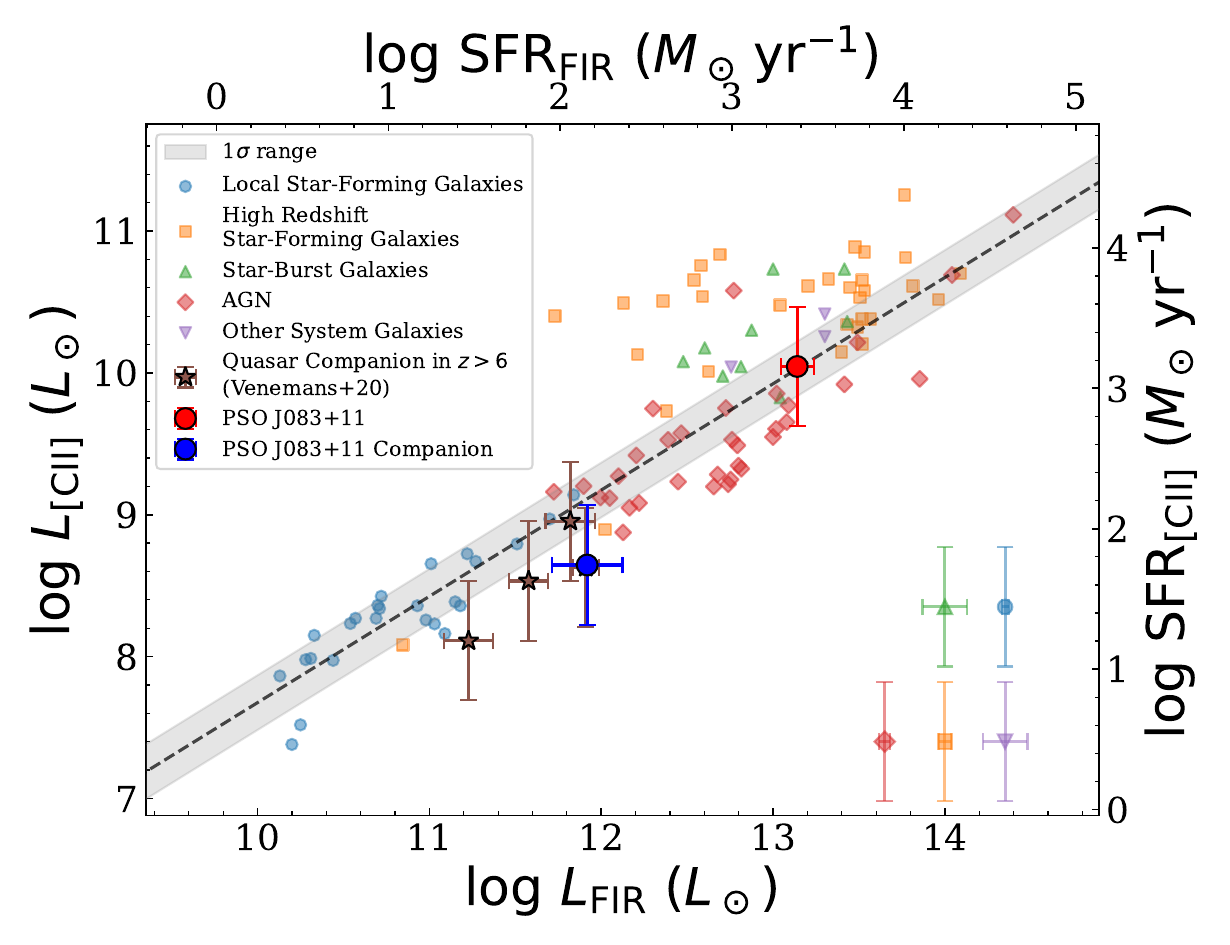}
		\caption{Correlation between [C\,\textsc{ii}] and FIR luminosities for the sample in Table~\ref{tab:galaxy_sample_extended}. The dashed line shows the best-fit relation with 1$\sigma$ dispersion (gray band). PSO J083+11 (large red circle) and its companion (large blue circle) lie within this scatter, consistent with normal star-forming galaxy behavior. The right axis indicates SFR derived from [C\,\textsc{ii}] luminosity using the \citet{DeLooze2014} calibration.}
		\label{fig:Hasil_Akhir}
	\end{figure}
	
	Figure~\ref{fig:cii_deficit} shows contrasting behaviors between the two diagnostic projections. In the $L_{\text{[C\,II]}}/L_{\text{FIR}}$ versus $L_{\text{FIR}}$ plane (left panel), both PSO J083+11 and its companion are located within the parameter space occupied by the broader galaxy population. The quasar positions near the transition region between high-redshift star-forming galaxies and starburst systems. The companion galaxy occupies a region populated by other $z > 6$ quasar companions from \citet{BPVenemans2020}, forming a coherent group with similar luminosity ratios. Comparing to the local star-forming galaxy population, both PSO J083+11 and its companion show somewhat lower luminosity ratios. However, this offset is consistent with the general trend observed for high-redshift star-forming galaxies and starbursts \citep{Stacey2010, Gullberg2015, Carniani2018}. Several AGN in the sample show significantly lower ratios, particularly at high FIR luminosities \citep{Stacey2010, Wagg2010, BPVenemans2020}. This placement indicates no significant [C\,\textsc{ii}] deficit when evaluated against total FIR luminosity, consistent with stellar heating-dominated ISM conditions \citep{Carniani2018, Schaerer2020, Gruppioni2020}.
	
	The $L_{\text{[C\,II]}}/L_{\text{FIR}}$ versus $\Sigma_{\text{FIR}}$ diagnostic reveals distinct behaviors for the two systems. The companion galaxy lies below both the \citet{Lutz2016} model and the \citet{DiazSantos2017} model. This position places the companion in a region also occupied by several other $z > 6$ quasar companions from \citet{BPVenemans2020}, suggesting that this may represent a characteristic property of satellite galaxies in high-redshift quasar environments. Local ultra-luminous infrared galaxies (ULIRGs) and several high-redshift AGN populate similar parameter space, characterized by compact FIR emission and relatively suppressed [C\,\textsc{ii}] luminosity for a given surface brightness. In contrast, PSO J083+11 occupies a distinct position, placing it above both theoretical models. The quasar is located in parameter space shared with several high-redshift star-forming galaxies and starburst systems, rather than clustering with the majority of AGN in the sample. The companion lies below the \citet{Lutz2016} prediction, while the quasar shows minimal deviation from this model. The divergence between integrated luminosity and surface brightness diagnostics demonstrates that the companion exhibits a surface brightness-driven [C\,\textsc{ii}] deficit when spatial concentration is considered, though it remains consistent with normal star-forming systems in the global $L_{\text{[C\,II]}}/L_{\text{FIR}}$ ratio.
	
	Figure~\ref{fig:Hasil_Akhir} displays the correlation between [C\,\textsc{ii}] and FIR luminosities for the full comparison sample. The dashed line represents the best-fit linear regression to the combined population, with a 1$\sigma$ dispersion indicated by the gray shaded region. The regression demonstrates a clear positive correlation between $L_{\text{[C\,II]}}$ and $L_{\text{FIR}}$ across more than four orders of magnitude in luminosity. PSO J083+11 is positioned within the 1$\sigma$ scatter of the correlation, lying among the high-redshift star-forming galaxies, starbursts, and several AGN that follow the general trend established by the broader population. The companion galaxy is positioned slightly below the regression line but within the 1$\sigma$ dispersion range when uncertainties are considered. This position is consistent with several other $z > 6$ quasar companions that show similar placements in this parameter space.
	
	The scatter in the $L_{\text{[C\,II]}}$ versus $L_{\text{FIR}}$ correlation is substantial at approximately 1~dex at a given FIR luminosity, reflecting the diversity of physical conditions across different galaxy populations \citep{Stacey2010, DeLooze2014, HerreraCamus2015, Carniani2018, Schaerer2020}. The spatial resolution of our ALMA observations corresponds to approximately 5.5~kpc at $z = 6.34$, comparable to studies of $z \sim 6$ quasar hosts by \citet{Decarli2018} and \citet{Wang2024}. At this resolution, both [C\,\textsc{ii}] and continuum emission appear spatially compact without extended structures, consistent with concentrated star formation in early massive galaxies.
	
	\subsection{Gas Dynamics and Distributions}
	
	\begin{figure*}
		\centering
		\includegraphics[width=2\columnwidth]{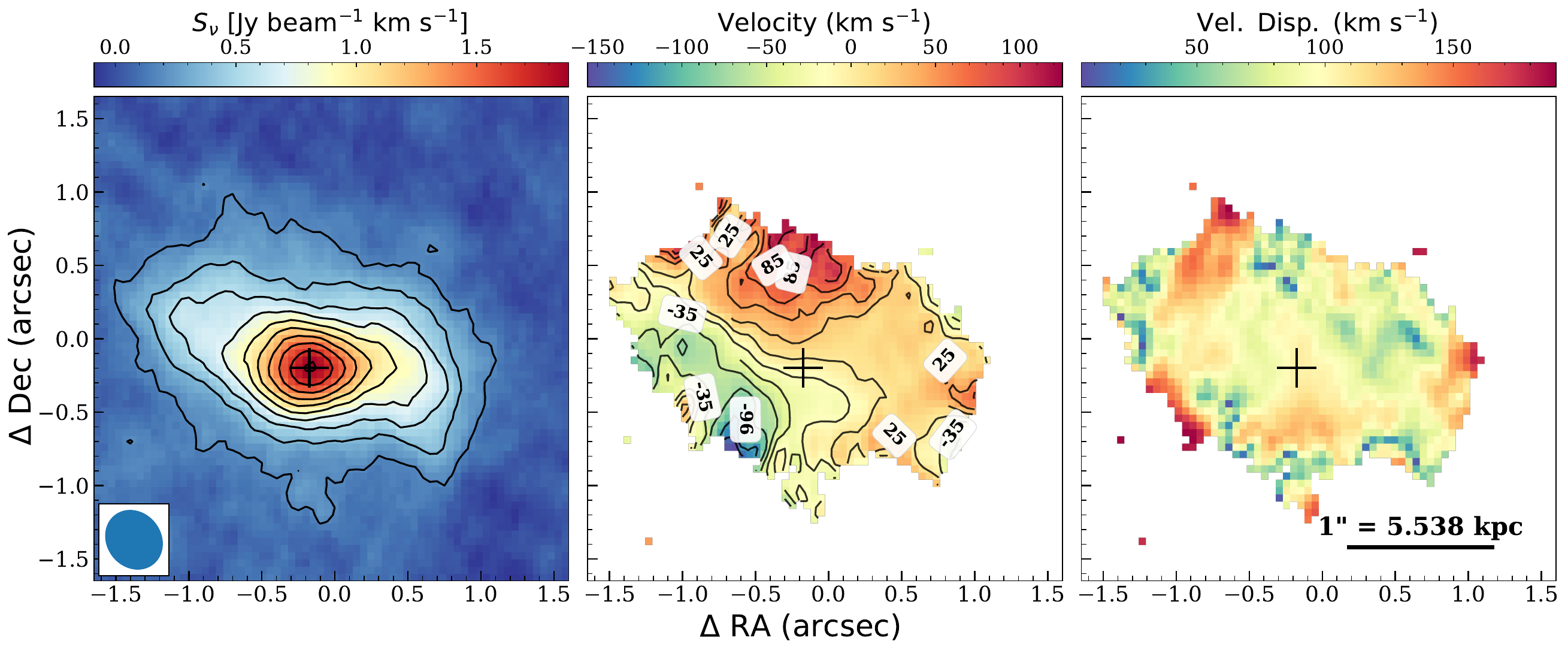}
		\caption{Quasar moment maps, each covering an area of $3\farcs312 \times 3\farcs312$ centered at $\alpha = 83.83700^\circ$ and $\delta = 11.84830^\circ$. (Left Panel) Quasar intensity map (moment 0). The lower left corner shows the beam size, and the lower right corner shows the scale conversion from $1''$ to kpc. The contour lines represent intensity levels [4, 12, 16, 20, 24, 28, 32, 36, 40]~$\times~\sigma$, measured from the outer region towards the center of the quasar. Here, $\sigma$ is the standard deviation of the background intensity calculated by excluding values above 3$\sigma$, resulting in $\sigma = 0.0465$ Jy beam$^{-1}$ km~s$^{-1}$. (Middle Panel) Velocity map (moment 1) of the quasar with maximum and minimum velocities of 124.982 and -155.836 km~s$^{-1}$.  The contours depict regions with different line-of-sight velocities, with numerical labels on the contours indicating the corresponding velocity values in km~s$^{-1}$. (Right Panel) Velocity dispersion map (moment 2) of the quasar with maximum and minimum values of 190.066 and 4.836 km~s$^{-1}$. The spatial range and resolution of this panel are identical to the intensity and velocity maps.}
		\label{fig:quasar_moment_maps}
	\end{figure*}
	
	\begin{figure*}
		\centering
		\includegraphics[width=2\columnwidth]{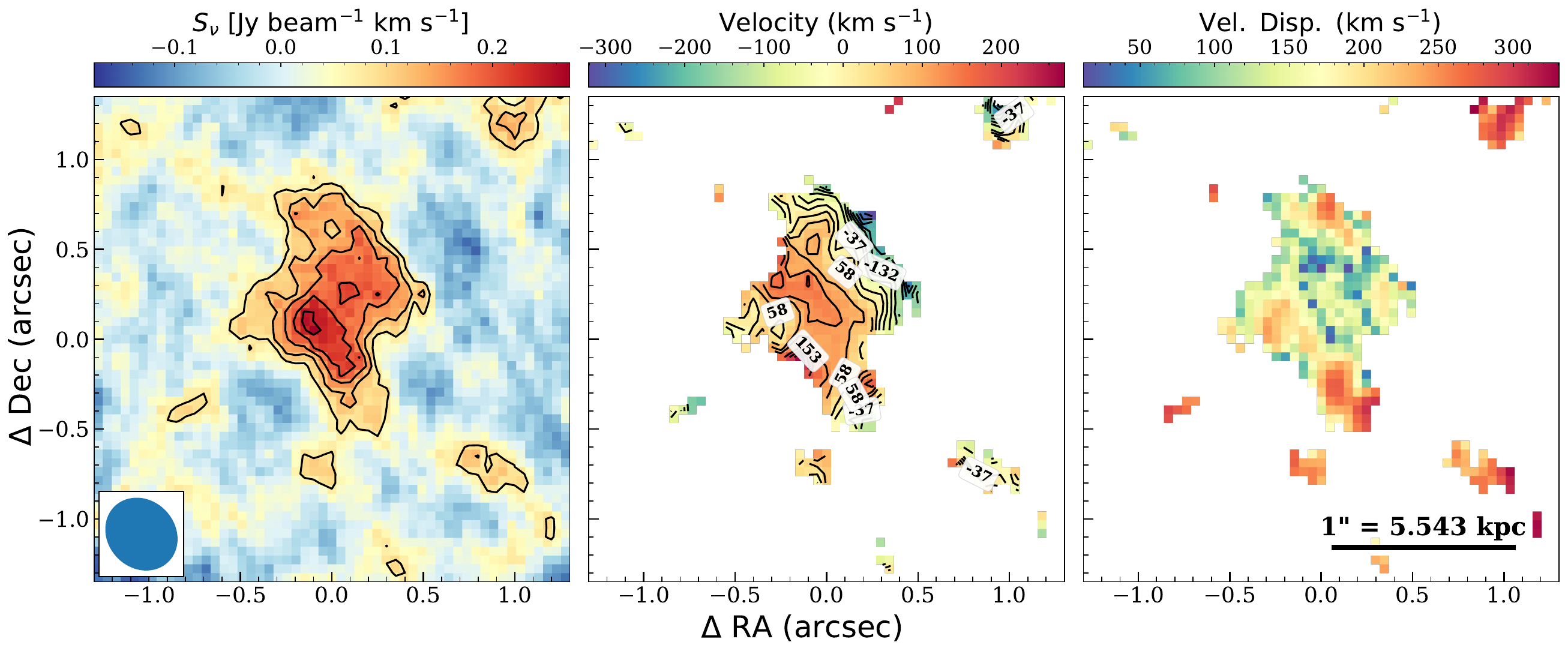}
		\caption{Moment maps of the companion galaxy, each covering an area of $2\farcs7 \times 2\farcs7$ centered at $\alpha = 83.8373^\circ$ and $\delta = 11.8474^\circ$. (Left panel) The intensity (moment 0) map of the companion galaxy. The contours represent intensity levels [2, 3, 4, 5, 6]~$\times~\sigma$. Here, $\sigma$ is the standard deviation of the background intensity calculated by excluding pixels with intensities above 2$\sigma$, resulting in $\sigma = 0.04099$~Jy~beam$^{-1}$~km~s$^{-1}$. (Middle Panel) The velocity (moment 1) map of the companion galaxy with maximum and minimum velocities of 279.511 and -321.217 km~s$^{-1}$. The contours indicate regions with different line-of-sight velocities, and the numerical values on the contours represent the corresponding velocities in km~s$^{-1}$. (Right Panel) The velocity dispersion (moment 2) map of the companion galaxy with maximum and minimum values of 303.939 and 12.059 km~s$^{-1}$. The spatial extent and resolution of this panel are consistent with those of the intensity and velocity maps.}
		\label{fig:companion_moment_maps}
	\end{figure*}
	
	\begin{figure*}
		\centering
		\includegraphics[width=0.98\columnwidth]{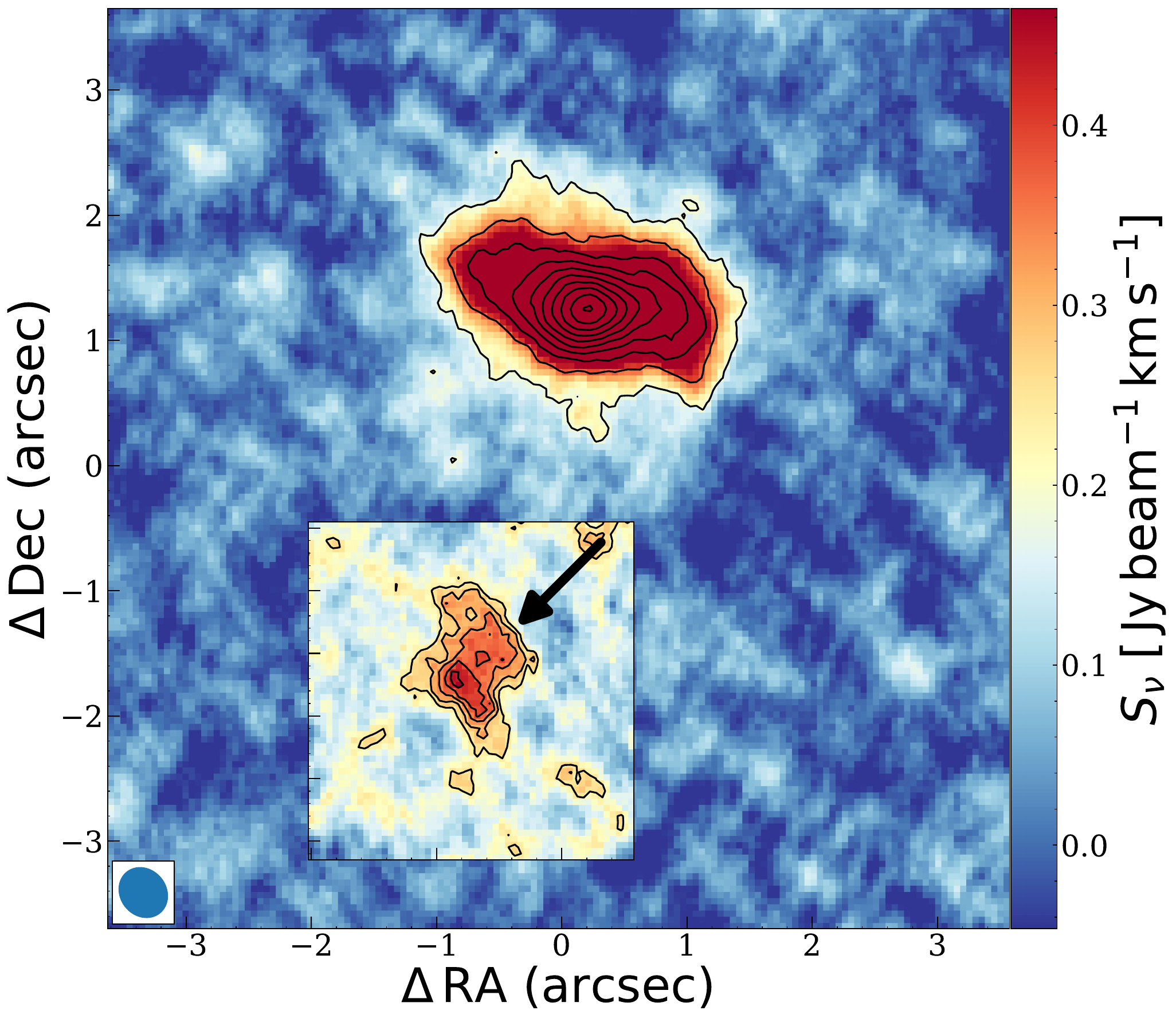}
		\includegraphics[width=1\columnwidth]{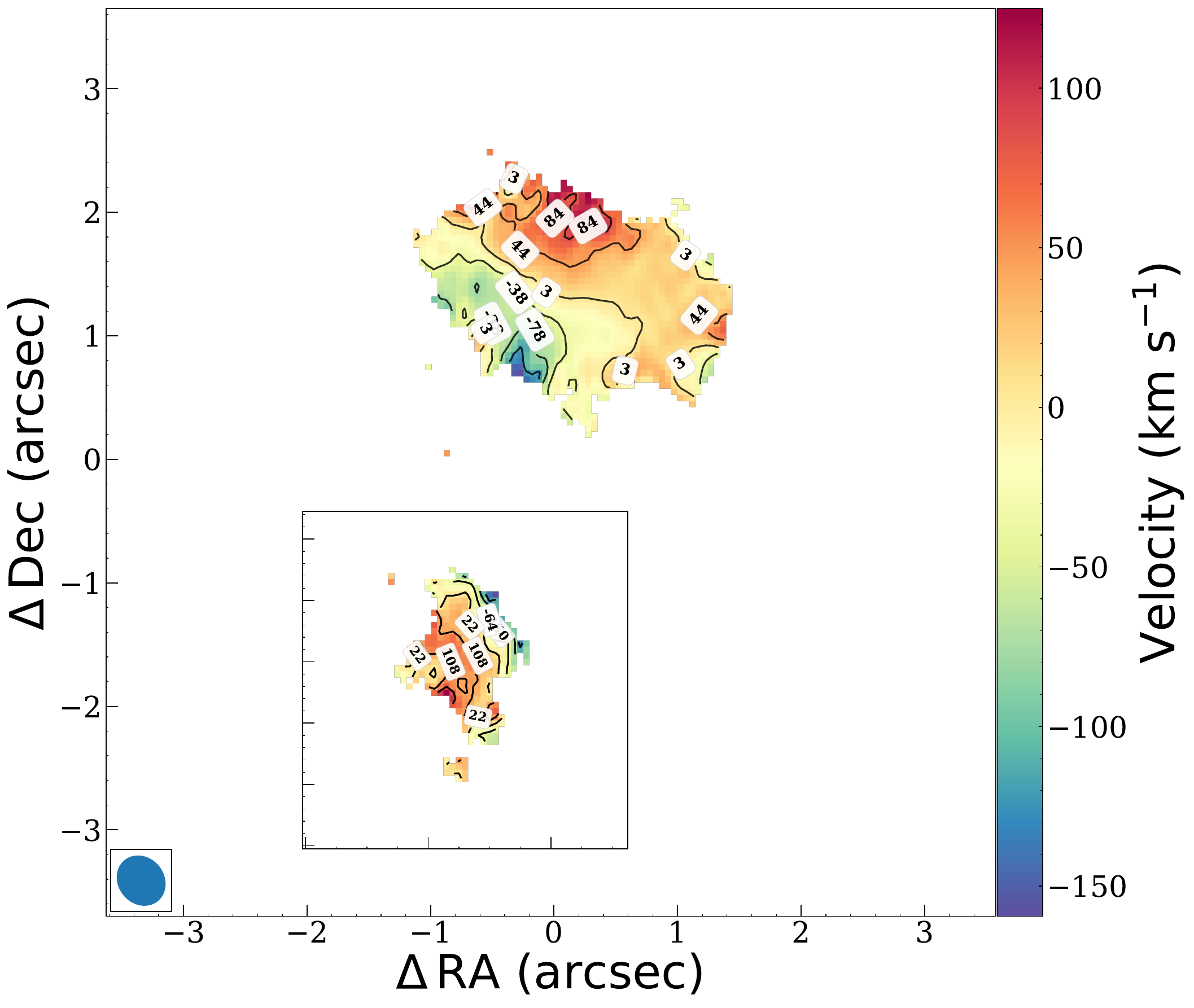}
		\includegraphics[width=1\columnwidth]{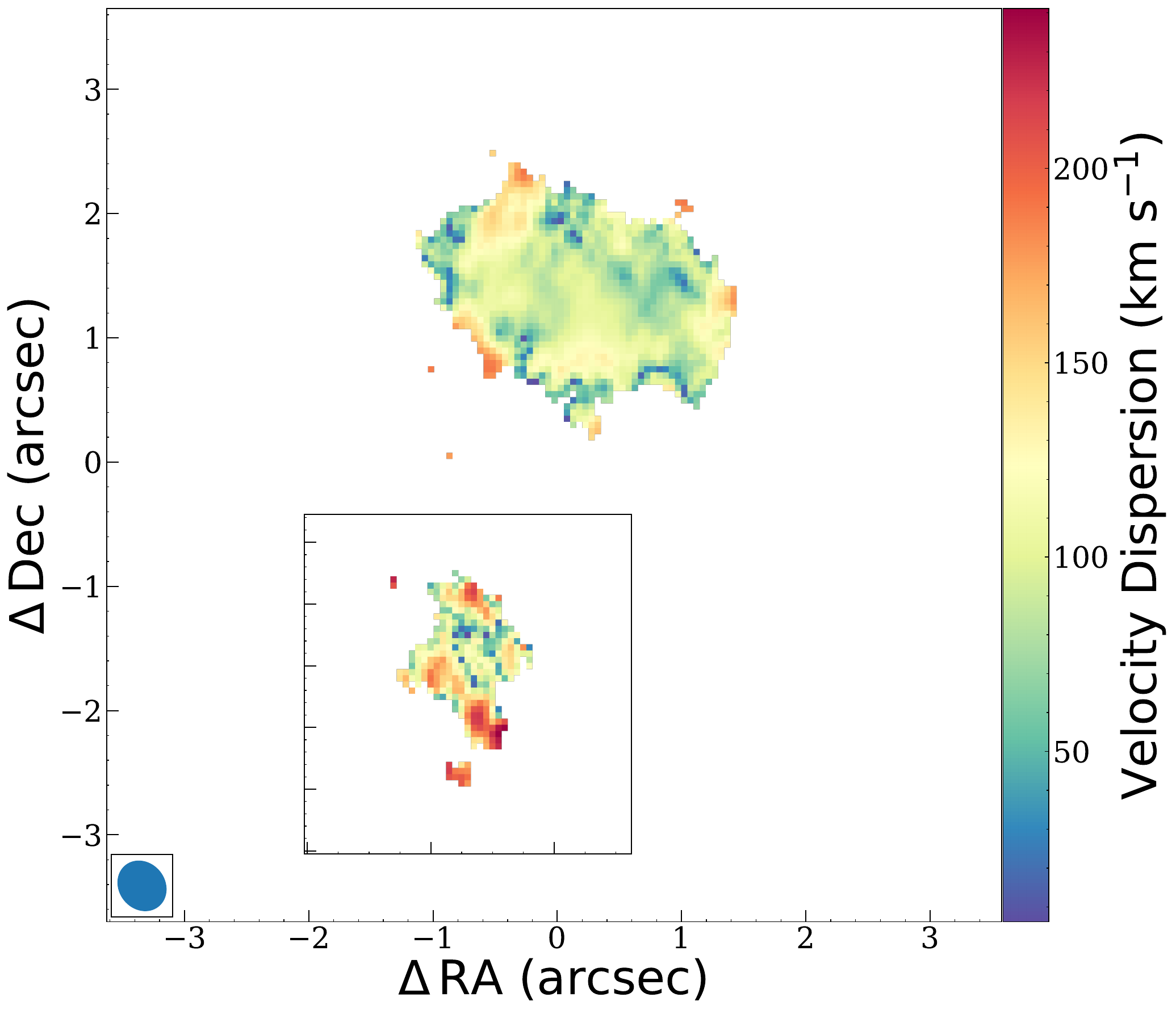}
		\caption{Combined moment maps of the quasar and its companion galaxy. All panels cover an area of $7\farcs336 \times 7\farcs336$. (Upper left panel) Combined intensity (moment 0) map. The contours and $\sigma$ value for the quasar are the same as in Figure~\ref{fig:quasar_moment_maps}, while the companion galaxy uses the same parameters as in Figure~\ref{fig:companion_moment_maps}. The arrow indicates a region of the companion galaxy that is likely influenced by AGN feedback. (Upper right panel) Combined velocity (moment 1) map with maximum and minimum values of 124.982 and -159.490 km~s$^{-1}$. For the quasar, pixels with intensities below 3$\sigma$ (using the $\sigma$ from Figure~\ref{fig:quasar_moment_maps}) are excluded. For the companion galaxy, a threshold of 2$\sigma$ is applied based on the $\sigma$ value from Figure~\ref{fig:companion_moment_maps}. (Lower Panel) Combined velocity dispersion (moment 2) map with maximum and minimum values of 241.134 and 6.160 km~s$^{-1}$, with the same spatial scale and resolution as upper panels.}
		\label{fig:combined_moment_maps}
	\end{figure*}
	
\begin{figure*}
		\centering
		\includegraphics[width=\columnwidth]{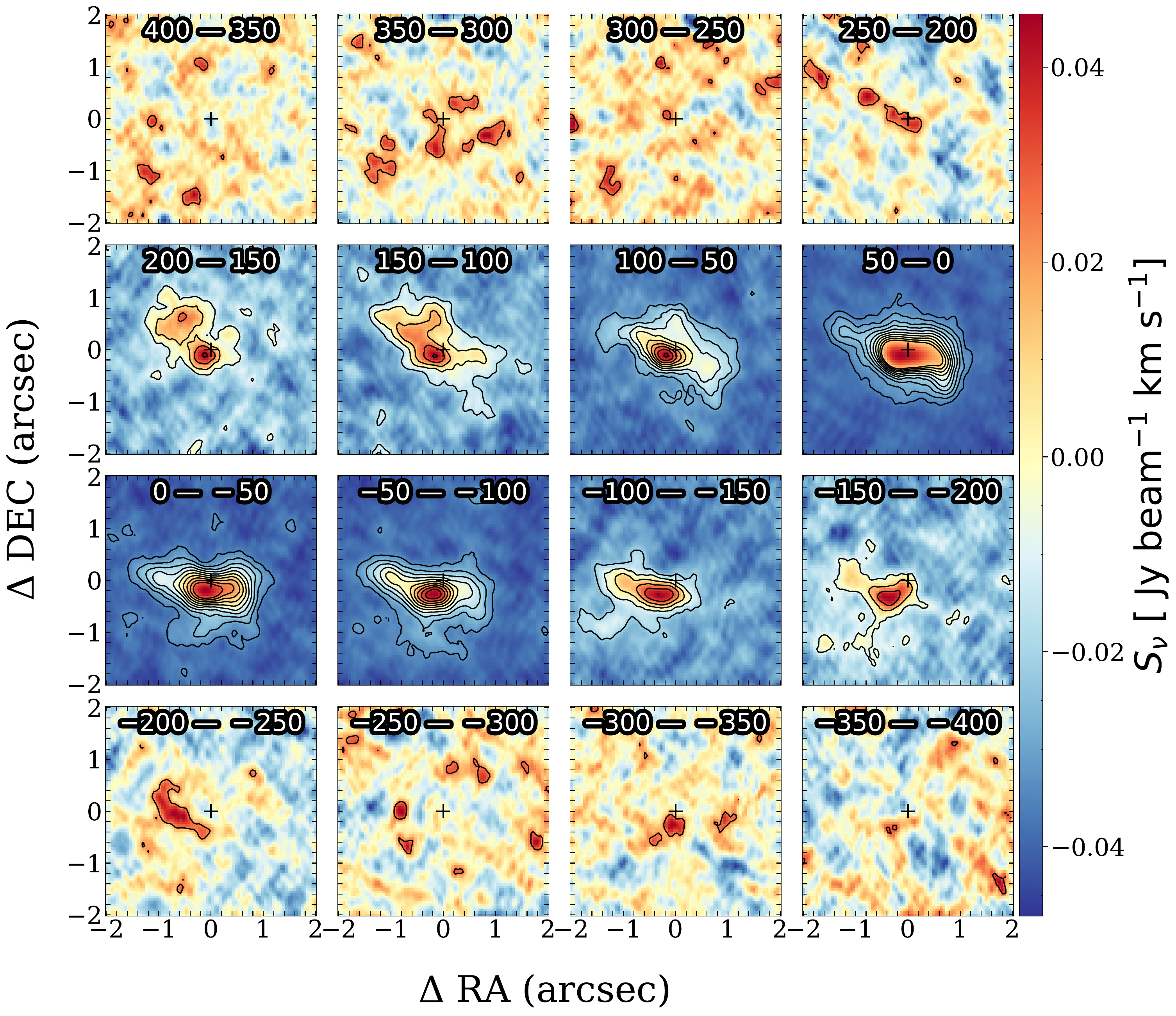}
		\includegraphics[width=\columnwidth]{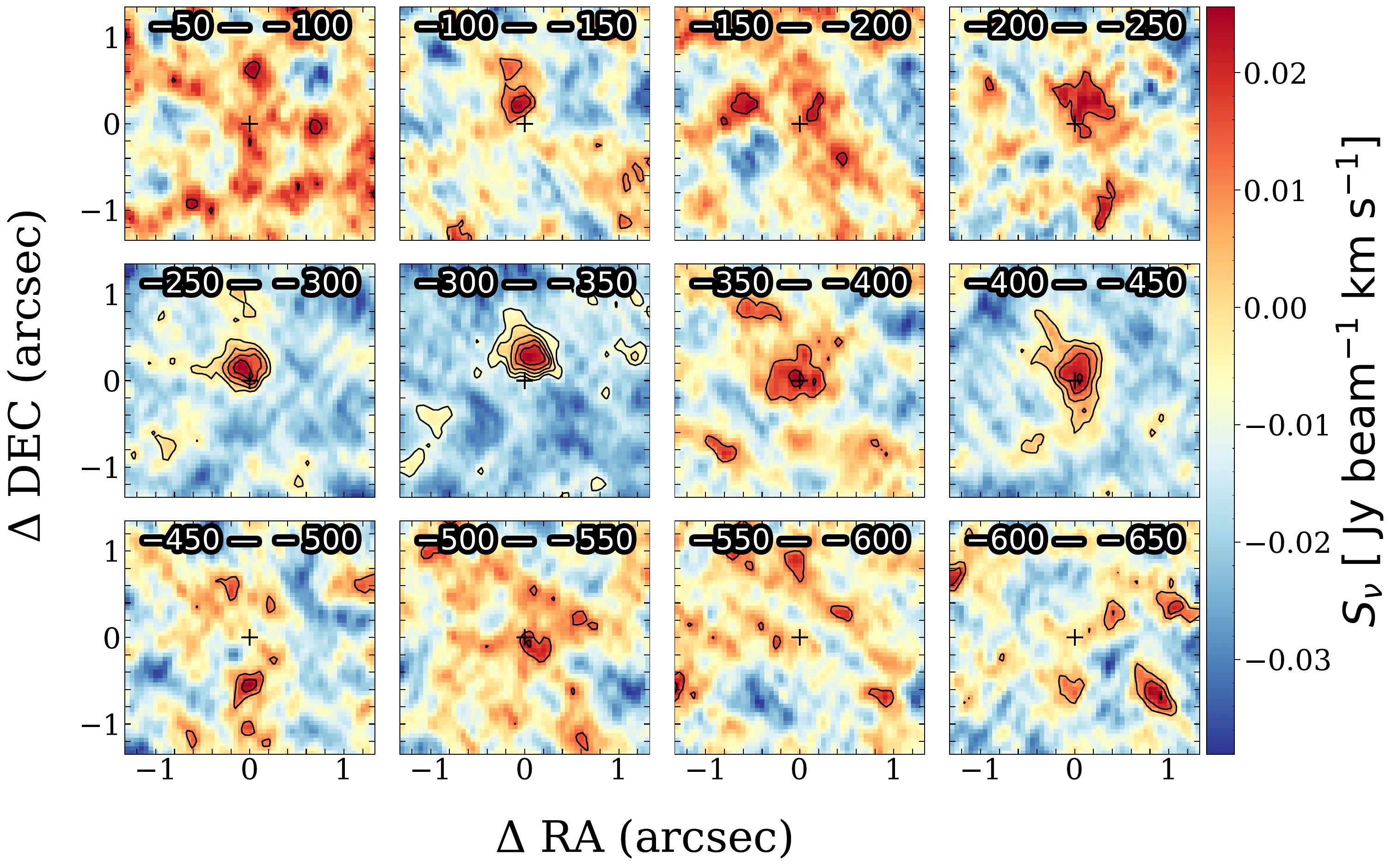}
		\caption{The quasar's moment channel map spans $4\farcs032 \times 4\farcs032$, 
			constructed from 50~km\,s$^{-1}$ velocity channels. The value shown at the top of each channel is given in units of ~km\,s$^{-1}$. Contours are drawn at $[3, 6, 9, 12, 15, 18, 21, 24, 27, 30] \times \sigma$, 
			where $\sigma$ varies per channel with values ranging from $0.0057$ to $0.0128$~Jy~beam$^{-1}$~km~s$^{-1}$ 
			(mean $\sigma = 0.0083$~Jy~beam$^{-1}$~km~s$^{-1}$). The central two rows (rows 2 and 3) highlight the dominant kinematic features 
			and spatial morphology. The velocity channel at $200$--$150$~km\,s$^{-1}$ exhibits a pronounced double-peak structure 
			exceeding 6$\sigma$ contours, indicating tidal features and merger signatures from ongoing dynamical interaction. 
			(Right panel)
			The companion galaxy's moment channel map covers a smaller area of $2\farcs657 \times 2\farcs700$, 
			constructed with the same 50~km\,s$^{-1}$ velocity channel resolution as the quasar. Contours are plotted at $[2, 3, 4, 5, 6] \times \sigma$, 
			with $\sigma$ ranging from $0.0072$ to $0.0131$~Jy~beam$^{-1}$~km~s$^{-1}$ 
			(mean $\sigma = 0.0096$~Jy~beam$^{-1}$~km~s$^{-1}$). The single middle row captures the kinematic signature, 
			revealing a more compact and organized gas structure.
		}
		\label{fig:moment_channel_maps}
	\end{figure*}

	The [C\,\textsc{ii}] moment maps and velocity channel maps, generated from the calibrated data cube using the \texttt{Spectral-Cube} package \citep{spectralcube2019}, reveal distinct cold gas properties and kinematic structures in PSO J083+11 and its companion galaxy (Figures~\ref{fig:quasar_moment_maps}--\ref{fig:moment_channel_maps}).
	
	The integrated intensity (moment~0) map of PSO J083+11 shows a compact, centrally concentrated [C\,\textsc{ii}] emission region with progressively nested contours, consistent with the spatially unresolved or marginally resolved morphologies typical of $z > 6$ quasar host galaxies at comparable angular resolutions \citep{Walter2004, Decarli2018, BPVenemans2020, Wang2024}. The absence of extended low-surface-brightness emission beyond the outermost contour is consistent with a system in an early phase of AGN activity. Theoretical models establish that radiation-driven winds are launched nearly immediately upon the onset of accretion, yet these winds must shock against the surrounding ISM and accumulate sufficient energy and momentum to drive galactic-scale flows on timescales of ${\sim}10^{5}$\,--\,$10^{6}$~yr \citep{ZubovasKing2012, KingPounds2015, Fabian2012}. Since typical AGN episodes last only ${\sim}10^{5}$~yr \citep{Schawinski2015}, many luminous quasars may be observed before large-scale feedback has propagated to kpc scales \citep{Costa2014, GaborBournaud2014}. The companion galaxy's moment~0 map shows more diffuse, lower-surface-brightness emission with a regular contour morphology free from elongation, arc-like structures, or secondary components that would indicate morphological disruption \citep{Carniani2018, Neeleman2019}. The combined map reveals two distinct emission concentrations without a detectable connecting bridge at the sensitivity of the present observations. A localized surface brightness enhancement on the northeastern edge of the companion galaxy reaches ${\sim}3\sigma$ significance and is spatially distinct from the main body of the companion emission.
	
	The intensity-weighted velocity field (moment~1) of PSO J083+11 displays a coherent, smoothly organized gradient whose regularity is characteristic of ordered rotation in a disc galaxy \citep{Neeleman2019, Pensabene2020}. The companion galaxy exhibits a broader velocity range across its emission region, consistent with a steep rotation curve or an edge-on geometric projection rather than kinematic disturbance \citep{Carniani2018, Pensabene2020}. Neither system shows the highly chaotic or asymmetric velocity patterns associated with powerful AGN-driven outflows or recent major mergers \citep{Swinbank2015, Neeleman2017, Carniani2018}. In the combined velocity map, the companion's emission appears at systematically more negative velocities relative to the quasar, consistent with the spectroscopic redshift offset of $\Delta z = 0.0092$, corresponding to a line-of-sight velocity separation of approximately $-376$~km~s$^{-1}$, confirming the two systems as kinematically distinct \citep{Carniani2013, Decarli2018}.
	
	The velocity dispersion (moment~2) maps show elevated values concentrated toward the central regions of both systems, partly attributable to beam smearing over unresolved velocity gradients \citep{Law2009, Neeleman2019}. The peak dispersions in both galaxies fall well below the extreme values ($\sigma_{v} \gtrsim 500$~km~s$^{-1}$) that characterize systems with powerful AGN-driven outflows \citep{Cicone2015, Butler2023, Tripodi2024}, and are consistent with the range measured in rotation-dominated $z > 6$ quasar hosts \citep{Neeleman2021, BPVenemans2020, Wang2024}. The combined dispersion map shows no interconnected high-$\sigma_{v}$ region bridging the two systems, which would otherwise be expected in the presence of large-scale interaction-driven turbulence \citep{Carniani2018, Decarli2018, Bischetti2024}.
	
	The $50$~km~s$^{-1}$ velocity channel maps, constructed following the approach of \citet{Banados2019}, provide additional kinematic diagnostics by revealing the spatial distribution of [C\,\textsc{ii}] emission at individual velocities independently of the integrated maps \citep{Carniani2013, Tripodi2024}. For the quasar, emission remains centrally concentrated and compact across most velocity channels, consistent with the moment~0 morphology. A notable exception occurs in the $+200$ to $+150$~km~s$^{-1}$ channel, where the emission displays a pronounced double-peaked spatial structure with both components exceeding $6\sigma$ significance and separated by approximately $1$~arcsec (${\sim}5.5$~kpc at $z = 6.34$). This double-peaked morphology is spatially resolved and distinct from a single centrally concentrated component, indicating that the [C\,\textsc{ii}] emission at this velocity originates from two spatially offset regions rather than from a single unresolved nucleus. The secondary component is displaced toward the upper-left direction relative to the primary emission peak, and weaker extensions in a similar direction persist in the adjacent $+150$ to $+100$~km~s$^{-1}$ channel and in several redshifted channels. The recurrence of this spatial offset across consecutive velocity channels indicates that it traces a coherent kinematic substructure rather than a noise artifact \citep{Banados2019, Decarli2018, Carniani2018}. Such velocity-localized spatial extensions have been interpreted in other $z \sim 6$ quasars as evidence of tidal bridges from gravitational interactions, distinct kinematic components within a rotating disc, or streaming motions associated with clumpy gas distributions \citep{Banados2019, Decarli2018, Carniani2018, Neeleman2019}. In the present case, the limited signal-to-noise ratio of the secondary component and the spatial resolution of these observations prevent a definitive classification of its physical origin, and higher-resolution follow-up observations would be required to distinguish between these scenarios. The companion channel maps present a contrasting picture, with emission remaining compact and centrally concentrated throughout its full velocity structure, showing no morphological asymmetries, spatial offsets, or secondary emission peaks reaching the $3\sigma$ significance threshold that would indicate gravitational perturbations or recent dynamical interactions \citep{Carniani2018, Neeleman2019, Novak2020}. Across all velocity channels, both systems maintain emission concentrated near their respective phase centers, and the smooth velocity gradients evident in the moment~1 maps together with the compact channel morphologies indicate that the dominant gas kinematics in both galaxies are consistent with rotation-dominated rather than outflow-disrupted configurations \citep{Cicone2014, Carniani2018, Molyneux2024}.
	
	\subsection{Distance of the Quasar--Companion System}
	
	The angular separation of $\theta = 3.2949$~arcsec between PSO J083+11 and its companion corresponds to a projected physical distance of $18.248 \pm 0.277$~kpc at $z = 6.34$, placing this configuration within the 5--60~kpc range typically observed in high-redshift quasar-companion systems \citep{Decarli2018, Neeleman2019, Mazzucchelli2019}. This position places the companion within the quasar proximity zone of $1.17 \pm 0.32$~pMpc \citep{Andika2020}, where theoretical models predict measurable AGN effects through ionization, gas heating, and photodissociation effects on the properties of nearby galaxies \citep{Eilers2017}.
	
	The derived projected separation is directly comparable with the statistical distribution reported by \citet{BPVenemans2020}, who surveyed 27 quasar host galaxies at $z \sim 6$ with ALMA at kiloparsec-scale resolution and found that 13 of 27 quasars have companion galaxies at projected separations spanning 3 to 90~kpc. Within this sample, the full range of configurations is represented, from close pairs such as PJ231$-$20 and PJ308$-$21 at ${\sim}8.5$ and ${\sim}14$~kpc respectively, to more separated companions such as J0842$+$1218 and J2100$-$1715 at up to ${\sim}50$~kpc \citep{BPVenemans2020, Pensabene2021}. The projected separation of $18.248 \pm 0.277$~kpc measured in PSO J083+11 places it firmly within the intermediate regime of this distribution, overlapping with the most commonly observed configurations in the \citet{BPVenemans2020} sample and consistent with the projected separation range found in the broader quasar-companion literature \citep{Decarli2018, Neeleman2019}.
	
	Despite being within the proximity zone, the companion galaxy retains a compact [C\,\textsc{ii}] morphology, regular kinematics, and spectral features indistinguishable from those of an isolated star-forming system, with no detectable signs of AGN radiative effects. This tranquility contrasts with expectations for galaxies in ionized regions, indicating that local geometric factors such as AGN collimation, dust filtering, or anisotropic emission patterns can significantly reduce the effectiveness of radiative coupling \citep{Farina2017, Decarli2017, fan2022quasarsintergalacticmediumcosmic}. Notably, several companions in the \citet{BPVenemans2020} sample at comparable or smaller separations also show no clear AGN-induced signatures in their [C\,\textsc{ii}] properties or kinematics, suggesting that proximity alone is insufficient to guarantee radiative feedback and that the projected physical distance, together with line-of-sight geometry and intervening gas column, collectively determines the observed ISM response \citep{BPVenemans2020, Decarli2018, Neeleman2019}.
	
	\section{Discussion}
	\label{sec:Discussion}
	
	Altogether, these three complementary diagnostics paint a consistent physical picture, in which neither the quasar host galaxy nor its companion experienced significant disturbance caused by the AGN at the time of observation. In the $L_{\rm [C\,II]}/L_{\rm FIR}$ versus $L_{\rm FIR}$ field, both PSO~J083+11 and its companion lie within the parameter space of the broader galaxy population, indicating no significant [C\,\textsc{ii}] deficit when evaluated against total FIR luminosity. This placement is naturally associated with stellar heating rather than AGN dominating the ISM conditions in both systems \citep{HerreraCamus2015, Carniani2018, Schaerer2020}. The companion falls below the theoretical models \citet{Lutz2016} and \citet{DiazSantos2017}, a position also shared by several other quasar companion with $z > 6$ in the \citet{BPVenemans2020} sample, suggesting that this surface brightness suppression reflects the compact FIR morphology of the satellite galaxy rather than external AGN radiation suppression \citep {DiazSantos2017, Decarli2018}. The quasar itself sits above both theoretical models in parameter space shared with high-redshift star-forming and starburst systems, consistent with stellar processes dominating its heating budget. In the direct $L_{\rm [C\,II]}$ versus $L_{\rm FIR}$ correlation, PSO~J083+11 lies within the $1\sigma$ scatter and the companion falls slightly below the regression line but remains within $1\sigma$ when its photometric uncertainties are propagated, confirming that both systems are consistent with the normal star-forming population across more than four orders of magnitude in luminosity \citep{DeLooze2014, Carniani2018, Schaerer2020}. The absence of a global [C\,\textsc{ii}] deficit and the tight spatial coincidence between [C\,\textsc{ii}] and FIR emission weigh strongly against scenarios in which the quasar clears or radiatively displaces gas in the companion's ISM, in agreement with the undisturbed companions studied by \citet{Pensabene2021} and \citet{Connor2020}. We therefore conclude that the AGN luminosity of PSO~J083+11 has not been sufficient to perturb the circumgalactic medium (CGM) at the position of the companion to a degree detectable in the present data, and that the proximity zone geometry alone does not guarantee observable AGN-driven effects \citep{Andika2020, Eilers2017}.
	
	One of the more striking aspects of this system is that its high SFR appears to be self-regulated rather than environmentally triggered. Rather than invoking the quasar as the driver of star-forming activity in either galaxy, the substantial cold gas reservoirs observed in both systems are more naturally attributed to gravity-driven disc evolution and cosmological gas accretion along large-scale filaments, processes expected to be highly efficient at $z > 6$ \citep{Wang2013, Decarli2017, Hodge2020}. At these epochs, the high gas fractions typical of the ISM, often exceeding 50\% of the dynamical mass, render galaxies susceptible to gravitational instabilities that fragment discs into massive star-forming clumps without requiring any external stimulus \citep{Haiman2006, Santini2009, Fensch2017}. Gas replenishment from the cosmic web can sustain SFR of this magnitude on timescales far exceeding any single AGN episode, decoupling the star formation history of the companion from the episodic behavior of the nearby quasar \citep{Narayanan2017}.
	
	One of the most compelling interpretations of the absence of large-scale AGN feedback is that PSO J083+11 is currently in a pre-outflow accretion phase. This is not a scenario requiring unusual conditions, but rather a natural consequence of the temporal hierarchy inherent in AGN feedback physics, in which radiation-driven winds are launched from the accretion disc on timescales far shorter than those required for the resulting shocks to propagate through the dense circumnuclear gas and break out into the CGM \citep{ZubovasKing2012, KingPounds2015, Fabian2012}. Numerical simulations by \citet{Costa2014} illustrate precisely this kind of buried configuration, in which the central black hole accretes vigorously and the system appears as a luminous quasar, yet the outflow remains confined within the inner few hundred parsecs and leaves no imprint at the kiloparsec-scale separations probed here. The observations presented in this work are entirely consistent with the early detection of PSO J083+11, and the observed quasar luminosity confirms that gas accretion onto the central SMBH is ongoing. In contrast, the unperturbed CGM at the position of the companion and the absence of any kinematic or morphological AGN signatures confirm that no large-scale outflow has yet been established \citep{Schawinski2015, GaborBournaud2014}. Future higher-resolution and higher-sensitivity observations could, in principle, detect nuclear-scale wind signatures or compact jet structures that would confirm this pre-outflow interpretation.
	
	The kinematic regularity observed across all moment maps is consistent with this picture and further discourages any interpretation of this system as the product of a recent major merger. The coherent velocity gradients and centrally concentrated velocity dispersions well below $\sigma_{v} \gtrsim 500$~km~s$^{-1}$ in both systems are characteristic of dynamically relaxed, rotationally supported discs rather than post-merger configurations, which typically exhibit disturbed and asymmetric kinematic fields for several dynamical times after coalescence \citep{Neeleman2019, Pensabene2020}. The only departure from this regularity is the double-peaked spatial structure detected in the quasar's $+200$ to $+150$~km~s$^{-1}$ velocity channel, where a secondary $[C\,\textsc{ii}]$-emitting component appears at ${\sim}1$~arcsec offset and persists into adjacent channels, constituting a coherent kinematic substructure that could represent a tidal bridge, a distinct rotating disc component, or gas streaming along clumpy structures \citep{Banados2019, Decarli2018, Carniani2018}. Similarly, the localized ${\sim}3\sigma$ surface brightness enhancement on the northeastern edge of the companion's moment~0 map may trace a minor interaction feature, but its low significance and spatial isolation from the main emission body prevent any firm conclusion about its origin. Considered together, these marginal features do not alter the dominant picture of two kinematically regular systems, but they do motivate follow-up observations at higher angular resolution to determine whether low-level tidal interaction or secular disc structures are responsible. This overall kinematic regularity places PSO~J083+11 in clear contrast with objects in the HYPERION quasar sample such as SDSS~J0100$+$2802 at $z \simeq 6.3$, which displays a prominent broad blueshifted [C\,\textsc{ii}] component interpreted as a large-scale gaseous outflow \citep{Tripodi2024}, and with merging systems at $z \sim 6.2$ displaying satellite accretion-driven morphological asymmetries \citep{Decarli2024}. The secular evolutionary pathway implied here challenges the paradigm that extreme SFR in high-redshift quasar environments are predominantly merger-driven \citep{Maiolino2003, Shangguan2020, Molina2023}.
	
	The projected physical separation of $18.248 \pm 0.277$~kpc places PSO~J083+11 and its companion in the intermediate regime of the \citet{BPVenemans2020} distribution, which spans 3 to 90~kpc across 13 detected quasar-companion pairs, bracketed on the close end by systems such as PJ231$-$20 and PJ308$-$21 at ${\sim}8.5$ and ${\sim}14$~kpc and on the wide end by J0842$+$1218 and J2100$-$1715 at ${\sim}50$~kpc \citep{BPVenemans2020, Pensabene2021}. The fact that several companions in this sample at comparable or even smaller projected separations also retain undisturbed [C\,\textsc{ii}] kinematics and show no AGN-induced spectroscopic signatures argues that the survival of dynamically quiet companions near luminous quasars is not anomalous but rather reflects the stochastic and anisotropic nature of AGN feedback at these epochs \citep{Fabian2012, KingPounds2015}. The apparent tension between studies reporting galaxy overdensities near $z \sim 6$ quasars \citep{McGreer2014, Zana2023} and those finding no field enhancement \citep{Yue2019} likely reflects this same diversity, with the observable outcome depending sensitively on AGN radiation cone orientation, evolutionary timing of the quasar episode, and local shielding by dust or gas columns along the line of sight \citep{Neeleman2019}.
	
	Viewed against the broader landscape of known quasar-companion systems, PSO~J083+11 and its companion most closely resemble the compact, internally regulated star-forming galaxies documented in \citet{Mazzucchelli2019} and \citet{Bischetti2024}, rather than feedback-affected systems such as the companion of SDSS~J0927$+$2943 \citep{Decarli2010}. The compactness of the companion's [C\,\textsc{ii}] emission and the regularity of its FIR morphology strongly disfavor any model in which the AGN has significantly reshaped the CGM at the companion's location. Instead, this system supports the broader conclusion that AGN feedback during the reionization era was neither spatially uniform nor temporally continuous, and that the coevolution of SMBHs and their host galaxies at $z > 6$ proceeded through phases of dynamically quiet growth punctuated by episodic feedback events \citep{Arp1981, Decarli2010, Shen2023}.

	\section{Conclusion}
	\label{sec:Conclusion}
	
	Utilizing high-resolution ALMA images at the [C\,\textsc{ii}] $158~\mu$m line and FIR continuum, we analysed the quasar-companion system PSO~J083+11 at $z = 6.34$ to assess whether bright quasars in the early universe significantly altered star formation or the CGM conditions of nearby galaxies through AGN feedback.
	
	Our results do not reveal convincing evidence for such an influence. The host galaxy of the quasar has $L_{\rm [C\,II]} = (1.110 \pm 0.065) \times 10^{10}~\mathrm{L_{\odot}}$ and $L_{\rm FIR} = (1.382 \pm 0.307) \times 10^{13}~\mathrm{L_{\odot}}$, corresponding to $\mathrm{SFR_{[C\,II]}} = 544$--$3764~\mathrm{M_{\odot}~yr^{-1}}$ and $\mathrm{SFR_{FIR}} = 1861$--$2932~\mathrm{M_{\odot}~yr^{-1}}$. The companion galaxy has a lower luminosity, with $L_{\rm [C\,II]} = (4.442 \pm 1.542) \times 10^{8}~\mathrm{L_{\odot}}$ and $L_{\rm FIR} = (8.283 \pm 3.920) \times 10^{11}~\mathrm{L_{\odot}}$, resulting in $\mathrm{SFR_{[C\,II]}} = 21$--$145~\mathrm{M_{\odot}~yr^{-1}}$ and $\mathrm{SFR_{FIR}} = 76$--$211~\mathrm{M_{\odot}~yr^{-1}}$. The peak velocity dispersion is $190.066~\mathrm{km~s^{-1}}$ for the quasar and $303.939~\mathrm{km~s^{-1}}$ for the companion galaxy, both consistent with rotationally supported discs.
	
	Across all three diagnostic projections, both galaxies are consistent with normal star-forming systems. In the $L_{\rm [C\,II]}/L_{\rm FIR}$ versus $L_{\rm FIR}$ plane, both systems fall within the parameter space of the broader galaxy population with no global [C\,\textsc{ii}] deficit. In the $L_{\rm [C\,II]}/L_{\rm FIR}$ versus $\Sigma_{\rm FIR}$ plane, the companion lies below the \citet{Lutz2016} and \citet{DiazSantos2017} models. This position better described as a surface brightness effect from its compact FIR morphology than as AGN-induced suppression, while the quasar sits above both models among high-redshift star-forming systems. In the direct $L_{\rm [C\,II]}$ versus $L_{\rm FIR}$ correlation, both systems lie within the $1\sigma$ scatter of the best-fit relation. The moment maps show compact, centrally peaked emission and coherent velocity gradients in both systems, with peak dispersions well below the $\sigma_{v} \gtrsim 500$~km~s$^{-1}$ threshold associated with powerful AGN-driven outflows. A double-peaked kinematic substructure is detected in the quasar's $+200$ to $+150$~km~s$^{-1}$ velocity channel and a marginal ${\sim}3\sigma$ surface brightness enhancement appears on the northeastern edge of the companion's moment~0 map, but neither feature shows sufficient significance or spatial extent to confirm tidal interaction or feedback-driven morphological disruption.
	
	Although the projected distance of $18.248 \pm 0.277$~kpc, which places PSO~J083+11 in the intermediate regime of the \citet{BPVenemans2020} companion distribution, situates the companion within the quasar proximity zone ($1.17 \pm 0.32$~pMpc), we find no evidence that the quasar has influenced the gas conditions or star formation in the companion through its CGM. We interpret this as a manifestation of a pre-outflow accretion phase, in which the gas is actively accreting onto the central SMBH as evidenced by the observed quasar luminosity. However, the AGN-driven wind has not yet propagated beyond the nuclear region to imprint large-scale mechanical or radiative signatures on the CGM at the companion's location \citep{Costa2014, KingPounds2015, Schawinski2015}. This interpretation is grounded in the theoretical expectation that outflow breakout timescales can exceed the duration of individual AGN episodes, making it entirely plausible for luminous quasars to be observed in dynamically quiet configurations before feedback becomes the dominant influence on their large-scale environment \citep{ZubovasKing2012, GaborBournaud2014, Fabian2012}.
	
	We conclude that at $z > 6$, black hole accretion and galaxy growth can proceed simultaneously in a dynamically quiet configuration without strong coupling between AGN activity and the surrounding CGM. The evolution of PSO~J083+11 and its companion is most naturally governed by internal processes including gradual disc growth, gravitational instability, and gas infall from the cosmic web, rather than by AGN-driven feedback propagating through the CGM. This system adds to the growing body of evidence that the coevolution of SMBHs and their host galaxies during the reionization era was episodic and anisotropic, and that the spatial reach of AGN feedback at these epochs was more limited than classical models have assumed.
	
	\section*{Acknowledgements}
	
	We gratefully acknowledge the financial support provided by the Faculty of Mathematics and Natural Sciences, Bandung Institute of Technology, through the Penelitian, Pengabdian Masyarakat, dan Inovasi (PPMI) 2026 research grant. 
	
	This paper makes use of the following ALMA data: ADS/JAO.ALMA\#2019.1.01436.S. ALMA is a partnership of ESO (representing its member states), NSF (USA), and NINS (Japan), together with NRC (Canada), MOST and ASIAA (Taiwan), and KASI (Republic of Korea), in cooperation with the Republic of Chile. The Joint ALMA Observatory is operated by ESO, AUI/NRAO and NAOJ. 
	
	The authors thank the anonymous referee for their constructive comments that significantly improved the quality of this manuscript. We also express our sincere gratitude to Alin Hafizhah Adira for her diligent assistance in proofreading and refining this work.
	
	
	\section*{Data Availability}
	
	The ALMA data underlying this article are available in the ALMA Science Archive at \url{https://almascience.eso.org/aq/} (Project ID: 2019.1.01436.S).
	
	
	\bibliographystyle{mnras}
	\bibliography{example}
	
	
	
	
	\appendix

\section{Galaxy Dataset for AGN Feedback Analysis}

We present 105 galaxies spanning $z \approx 0.03$--$7.5$ (Table~\ref{tab:galaxy_sample_extended}):
25 local star-forming galaxies, 32 high-redshift star-forming systems, 33 AGN, 8 starbursts,
3 composite systems, and 4 quasar companions. All luminosities are in solar units ($L_\odot$);
$\log(L_{\text{[C\,\textsc{ii}]}}/L_{\text{FIR}})$ quantifies the [C\,\textsc{ii}] deficit,
and $\Sigma_{\text{FIR}}$ denotes FIR surface brightness in $L_\odot\,\text{kpc}^{-2}$.

\onecolumn

\setlength{\LTcapwidth}{\textwidth}

\begin{longtable}{@{}lcccccc@{}}

	\caption{Galaxy sample for AGN feedback analysis.
		References:
		[1]~\citet{Hughes2017},
		[2]~\citet{Brisbin2015},
		[3]~\citet{Stacey2010},
		[4]~\citet{Schaerer2015},
		[5]~\citet{Gullberg2015},
		[6]~\citet{Wagg2010},
		[7]~\citet{Iono2006},
		[8]~\citet{BPVenemans2020},
		[9]~\citet{Walter2012},
		[10]~\citet{Riechers2014},
		[11]~\citet{DeBreuck2011},
		[12]~\citet{Marsden2005},
		[13]~\citet{Hashimoto2019}.
		$^{a}$\,$\Sigma_{\text{FIR}}$ estimated via the size--luminosity
		relation of \citet{Lutz2016};
		$^{b}$\,direct measurement from \citet{Decarli2018}.}
	\label{tab:galaxy_sample_extended} \\
	\hline
	Galaxy &
	$z$ &
	$\log(L_{\text{[C\,\textsc{ii}]}})$ &
	$\log(L_{\text{FIR}})$ &
	$\log(L_{\text{[C\,\textsc{ii}]}}/L_{\text{FIR}})$ &
	$\log(\Sigma_{\text{FIR}})$ &
	Ref \\
	\hline
	\endfirsthead   
	
	\hline
	Galaxy &
	$z$ &
	$\log(L_{\text{[C\,\textsc{ii}]}})$ &
	$\log(L_{\text{FIR}})$ &
	$\log(L_{\text{[C\,\textsc{ii}]}}/L_{\text{FIR}})$ &
	$\log(\Sigma_{\text{FIR}})$ &
	Ref \\
	\hline
	\endhead        
	
	\hline
	\multicolumn{7}{r}{\textit{Continued on next page\ldots}} \\
	\endfoot        
	
	\hline
	\endlastfoot     
	
	\multicolumn{7}{c}{\textbf{Local Star-Forming Galaxies}} \\
	\hline
	G09.DR1.12  & 0.182 & $9.140 \pm 0.021$ & $11.840 \pm 0.020$ & $-2.700 \pm 0.029$ & $11.588 \pm 0.152^{a}$ & [1] \\
	G09.DR1.20  & 0.055 & $8.161 \pm 0.015$ & $11.090 \pm 0.010$ & $-2.929 \pm 0.018$ & $10.532 \pm 0.107^{a}$ & [1] \\
	G09.DR1.24  & 0.033 & $7.987 \pm 0.009$ & $10.310 \pm 0.020$ & $-2.323 \pm 0.022$ & $9.433  \pm 0.080^{a}$ & [1] \\
	G09.DR1.32  & 0.072 & $8.671 \pm 0.008$ & $11.270 \pm 0.010$ & $-2.599 \pm 0.013$ & $10.785 \pm 0.116^{a}$ & [1] \\
	G09.DR1.37  & 0.059 & $8.362 \pm 0.015$ & $10.700 \pm 0.020$ & $-2.338 \pm 0.025$ & $9.983  \pm 0.092^{a}$ & [1] \\
	G09.DR1.43  & 0.054 & $8.270 \pm 0.014$ & $10.570 \pm 0.020$ & $-2.300 \pm 0.024$ & $9.800  \pm 0.087^{a}$ & [1] \\
	G09.DR1.47  & 0.026 & $7.519 \pm 0.013$ & $10.250 \pm 0.010$ & $-2.731 \pm 0.017$ & $9.349  \pm 0.075^{a}$ & [1] \\
	G09.DR1.49  & 0.051 & $8.338 \pm 0.008$ & $10.710 \pm 0.010$ & $-2.372 \pm 0.013$ & $9.997  \pm 0.089^{a}$ & [1] \\
	G09.DR1.53  & 0.107 & $8.724 \pm 0.017$ & $11.220 \pm 0.010$ & $-2.496 \pm 0.020$ & $10.715 \pm 0.114^{a}$ & [1] \\
	G09.DR1.56  & 0.060 & $8.425 \pm 0.011$ & $10.720 \pm 0.020$ & $-2.295 \pm 0.023$ & $10.011 \pm 0.093^{a}$ & [1] \\
	G09.DR1.60  & 0.052 & $8.270 \pm 0.009$ & $10.690 \pm 0.020$ & $-2.420 \pm 0.022$ & $9.969  \pm 0.092^{a}$ & [1] \\
	G09.DR1.61  & 0.053 & $7.973 \pm 0.014$ & $10.440 \pm 0.020$ & $-2.467 \pm 0.024$ & $9.617  \pm 0.083^{a}$ & [1] \\
	G09.DR1.62  & 0.133 & $8.794 \pm 0.044$ & $11.510 \pm 0.020$ & $-2.716 \pm 0.048$ & $11.123 \pm 0.132^{a}$ & [1] \\
	G09.DR1.72  & 0.079 & $8.654 \pm 0.013$ & $11.010 \pm 0.030$ & $-2.356 \pm 0.033$ & $10.419 \pm 0.110^{a}$ & [1] \\
	G09.DR1.80  & 0.073 & $8.230 \pm 0.023$ & $11.030 \pm 0.010$ & $-2.800 \pm 0.025$ & $10.447 \pm 0.104^{a}$ & [1] \\
	G09.DR1.85  & 0.031 & $7.863 \pm 0.006$ & $10.130 \pm 0.030$ & $-2.267 \pm 0.031$ & $9.180  \pm 0.084^{a}$ & [1] \\
	G09.DR1.99  & 0.128 & $8.970 \pm 0.019$ & $11.700 \pm 0.010$ & $-2.730 \pm 0.021$ & $11.391 \pm 0.141^{a}$ & [1] \\
	G09.DR1.113 & 0.078 & $8.386 \pm 0.018$ & $11.150 \pm 0.010$ & $-2.764 \pm 0.020$ & $10.616 \pm 0.110^{a}$ & [1] \\
	G09.DR1.125 & 0.041 & $7.978 \pm 0.009$ & $10.280 \pm 0.030$ & $-2.302 \pm 0.031$ & $9.391  \pm 0.086^{a}$ & [1] \\
	G09.DR1.159 & 0.044 & $8.233 \pm 0.008$ & $10.540 \pm 0.020$ & $-2.307 \pm 0.021$ & $9.757  \pm 0.086^{a}$ & [1] \\
	G09.DR1.179 & 0.070 & $8.358 \pm 0.017$ & $11.180 \pm 0.010$ & $-2.822 \pm 0.020$ & $10.658 \pm 0.111^{a}$ & [1] \\
	G09.DR1.185 & 0.051 & $8.149 \pm 0.012$ & $10.330 \pm 0.030$ & $-2.181 \pm 0.032$ & $9.462  \pm 0.087^{a}$ & [1] \\
	G09.DR1.276 & 0.027 & $7.380 \pm 0.018$ & $10.200 \pm 0.010$ & $-2.820 \pm 0.021$ & $9.279  \pm 0.075^{a}$ & [1] \\
	G09.DR1.294 & 0.096 & $8.358 \pm 0.021$ & $10.930 \pm 0.040$ & $-2.572 \pm 0.045$ & $10.306 \pm 0.113^{a}$ & [1] \\
	G09.DR1.328 & 0.074 & $8.258 \pm 0.024$ & $10.980 \pm 0.010$ & $-2.722 \pm 0.026$ & $10.377 \pm 0.101^{a}$ & [1] \\
	\hline
	
	\multicolumn{7}{c}{\textbf{High-Redshift Star-Forming Galaxies}} \\
	\hline
	MIPS 22530                      & 1.950 & $10.836 \pm 0.101$ & $12.690 \pm 0.040$ & $-1.854 \pm 0.109$ & $12.785 \pm 0.212^{a}$ & [2] \\
	SWIRE3 J104343.93+571322.5      & 1.735 & $10.539 \pm 0.235$ & $12.590 \pm 0.150$ & $-2.051 \pm 0.279$ & $12.644 \pm 0.289^{a}$ & [2] \\
	SWIRE3 J104514.38+575708.8      & 1.780 & $10.129 \pm 0.200$ & $12.210 \pm 0.130$ & $-2.081 \pm 0.238$ & $12.109 \pm 0.252^{a}$ & [2] \\
	SWIRE3 J104632.93+563530.2      & 1.771 & $10.492 \pm 0.199$ & $12.130 \pm 0.130$ & $-1.638 \pm 0.238$ & $11.996 \pm 0.248^{a}$ & [2] \\
	SMM J030227.73+000653.5         & 1.408 & $10.756 \pm 0.092$ & $12.580 \pm 0.030$ & $-1.824 \pm 0.097$ & $12.630 \pm 0.201^{a}$ & [2] \\
	SWIRE4 J104427.52+584309.6      & 1.756 & $10.402 \pm 0.292$ & $11.730 \pm 0.190$ & $-1.328 \pm 0.348$ & $11.433 \pm 0.303^{a}$ & [2] \\
	SWIRE4 J104656.46+590235.5      & 1.854 & $10.506 \pm 0.239$ & $12.360 \pm 0.150$ & $-1.854 \pm 0.282$ & $12.320 \pm 0.279^{a}$ & [2] \\
	SDSS J120602.09+514229.5        & 1.998 & $10.654 \pm 0.105$ & $12.540 \pm 0.030$ & $-1.886 \pm 0.109$ & $12.573 \pm 0.199^{a}$ & [2] \\
	MACS J0451+0006.                & 2.013 & $8.083  \pm 0.043$ & $10.845 \pm 0.124$ & $-2.762 \pm 0.131$ & $10.187 \pm 0.198^{a}$ & [4] \\
	GRB 080207                      & 2.086 & $8.895  \pm 0.118$ & $12.021 \pm 0.054$ & $-3.126 \pm 0.130$ & $11.843 \pm 0.178^{a}$ & [13] \\
	SPT0551-50                      & 2.123 & $10.477 \pm 0.000$ & $13.041 \pm 0.039$ & $-2.564 \pm 0.039$ & $13.279 \pm 0.234^{a}$ & [5] \\
	SPT0512-59                      & 2.234 & $10.602 \pm 0.087$ & $13.447 \pm 0.031$ & $-2.845 \pm 0.092$ & $13.851 \pm 0.259^{a}$ & [5] \\
	SPT0538-50                      & 2.782 & $11.255 \pm 0.000$ & $13.763 \pm 0.022$ & $-2.508 \pm 0.022$ & $14.296 \pm 0.279^{a}$ & [5] \\
	SPT0103-45                      & 3.090 & $10.580 \pm 0.057$ & $13.531 \pm 0.026$ & $-2.952 \pm 0.063$ & $13.969 \pm 0.263^{a}$ & [5] \\
	SPT0550-53                      & 3.129 & $10.613 \pm 0.085$ & $13.204 \pm 0.027$ & $-2.591 \pm 0.089$ & $13.508 \pm 0.242^{a}$ & [5] \\
	SPT0529-54                      & 3.369 & $10.886 \pm 0.039$ & $13.477 \pm 0.029$ & $-2.591 \pm 0.049$ & $13.893 \pm 0.260^{a}$ & [5] \\
	SPT0532-50                      & 3.399 & $10.613 \pm 0.074$ & $13.813 \pm 0.027$ & $-3.200 \pm 0.079$ & $14.366 \pm 0.283^{a}$ & [5] \\
	SPT0300-46                      & 3.596 & $10.204 \pm 0.109$ & $13.519 \pm 0.026$ & $-3.314 \pm 0.112$ & $13.951 \pm 0.263^{a}$ & [5] \\
	SPT2147-50                      & 3.761 & $10.531 \pm 0.064$ & $13.505 \pm 0.027$ & $-2.974 \pm 0.069$ & $13.932 \pm 0.262^{a}$ & [5] \\
	SPT0418-47                      & 4.224 & $10.813 \pm 0.033$ & $13.771 \pm 0.029$ & $-2.958 \pm 0.045$ & $14.306 \pm 0.280^{a}$ & [5] \\
	SPT0113-46                      & 4.232 & $10.663 \pm 0.094$ & $13.322 \pm 0.021$ & $-2.659 \pm 0.097$ & $13.675 \pm 0.248^{a}$ & [5] \\
	SPT2311-54                      & 4.281 & $10.380 \pm 0.054$ & $13.519 \pm 0.039$ & $-3.138 \pm 0.067$ & $13.951 \pm 0.266^{a}$ & [5] \\
	SPT0345-47                      & 4.296 & $10.519 \pm 0.053$ & $13.964 \pm 0.038$ & $-3.445 \pm 0.065$ & $14.578 \pm 0.295^{a}$ & [5] \\
	SPT2103-60                      & 4.435 & $10.851 \pm 0.061$ & $13.531 \pm 0.026$ & $-2.680 \pm 0.066$ & $13.969 \pm 0.263^{a}$ & [5] \\
	SPT0441-46                      & 4.477 & $10.380 \pm 0.109$ & $13.568 \pm 0.023$ & $-3.188 \pm 0.111$ & $14.021 \pm 0.265^{a}$ & [5] \\
	SPT2146-55                      & 4.567 & $10.342 \pm 0.099$ & $13.431 \pm 0.048$ & $-3.089 \pm 0.110$ & $13.828 \pm 0.263^{a}$ & [5] \\
	SPT2132-58                      & 4.768 & $10.322 \pm 0.083$ & $13.491 \pm 0.042$ & $-3.169 \pm 0.093$ & $13.913 \pm 0.265^{a}$ & [5] \\
	SPT2319-55                      & 5.293 & $10.146 \pm 0.062$ & $13.398 \pm 0.035$ & $-3.252 \pm 0.071$ & $13.781 \pm 0.256^{a}$ & [5] \\
	SPT0346-52                      & 5.656 & $10.699 \pm 0.061$ & $14.090 \pm 0.018$ & $-3.391 \pm 0.063$ & $14.756 \pm 0.300^{a}$ & [5] \\
	SPT0243-49                      & 5.699 & $10.653 \pm 0.000$ & $13.519 \pm 0.039$ & $-2.865 \pm 0.039$ & $13.951 \pm 0.266^{a}$ & [5] \\
	LESS J033229.4$-$275619         & 4.753 & $10.009 \pm 0.064$ & $12.623 \pm 0.186$ & $-2.615 \pm 0.197$ & $12.691 \pm 0.329^{a}$ & [11] \\
	CL 1358+62-G1                   & 4.926 & $9.732  \pm 0.130$ & $12.380 \pm 0.217$ & $-2.648 \pm 0.253$ & $12.348 \pm 0.357^{a}$ & [12] \\
	\hline
	
	\multicolumn{7}{c}{\textbf{Active Galactic Nuclei}} \\
	\hline
	PKS 0215+015               & 1.720 & $10.690 \pm 0.130$ & $14.041 \pm 0.130$ & $-3.351 \pm 0.184$ & $14.687 \pm 0.348^{a}$ & [3] \\
	PG1206+459                 & 1.159 & $9.919  \pm 0.130$ & $13.415 \pm 0.130$ & $-3.496 \pm 0.184$ & $13.805 \pm 0.312^{a}$ & [3] \\
	PG1241+176                 & 1.273 & $10.580 \pm 0.217$ & $12.771 \pm 0.130$ & $-2.191 \pm 0.253$ & $12.898 \pm 0.278^{a}$ & [3] \\
	IRAS F22231--0512 (3C~446) & 1.403 & $11.114 \pm 0.217$ & $14.398 \pm 0.130$ & $-3.284 \pm 0.253$ & $15.189 \pm 0.369^{a}$ & [3] \\
	BRI 1335-0417              & 4.407 & $10.215 \pm 0.069$ & $13.491 \pm 0.066$ & $-3.277 \pm 0.095$ & $13.913 \pm 0.274^{a}$ & [6] \\
	BR 1202N                   & 4.691 & $9.653  \pm 0.068$ & $13.079 \pm 0.072$ & $-3.426 \pm 0.099$ & $13.332 \pm 0.252^{a}$ & [7] \\
	J0129-0035                 & 5.779 & $9.283  \pm 0.016$ & $12.678 \pm 0.010$ & $-3.394 \pm 0.019$ & $12.767 \pm 0.204^{b}$ & [8] \\
	J1044-0125                 & 5.785 & $9.215  \pm 0.056$ & $12.739 \pm 0.017$ & $-3.524 \pm 0.058$ & $12.853 \pm 0.209^{b}$ & [8] \\
	P007+04                    & 6.002 & $9.199  \pm 0.025$ & $12.655 \pm 0.011$ & $-3.456 \pm 0.027$ & $11.800 \pm 0.065^{b}$ & [8] \\
	P009-10                    & 6.004 & $9.958  \pm 0.032$ & $13.853 \pm 0.004$ & $-3.895 \pm 0.032$ & $11.550 \pm 0.045^{b}$ & [8] \\
	J0100+2802                 & 6.327 & $9.575  \pm 0.020$ & $12.465 \pm 0.027$ & $-2.890 \pm 0.033$ & $12.468 \pm 0.193^{b}$ & [8] \\
	J0109-3047                 & 6.790 & $9.272  \pm 0.037$ & $12.100 \pm 0.065$ & $-2.829 \pm 0.075$ & $11.954 \pm 0.190^{b}$ & [8] \\
	J025-33                    & 6.337 & $9.752  \pm 0.017$ & $12.725 \pm 0.020$ & $-2.973 \pm 0.026$ & $12.834 \pm 0.208^{b}$ & [8] \\
	P036+03                    & 6.540 & $9.529  \pm 0.012$ & $12.761 \pm 0.009$ & $-3.232 \pm 0.015$ & $12.885 \pm 0.209^{b}$ & [8] \\
	J0305-3150                 & 6.614 & $9.771  \pm 0.026$ & $13.090 \pm 0.016$ & $-3.319 \pm 0.031$ & $13.348 \pm 0.232^{b}$ & [8] \\
	P065-26                    & 6.187 & $9.233  \pm 0.043$ & $12.447 \pm 0.036$ & $-3.214 \pm 0.056$ & $11.150 \pm 0.055^{b}$ & [8] \\
	J0842+1218                 & 6.075 & $8.875  \pm 0.058$ & $12.127 \pm 0.036$ & $-3.252 \pm 0.068$ & $10.250 \pm 0.145^{b}$ & [8] \\
	J1048-0109                 & 6.676 & $9.324  \pm 0.031$ & $12.816 \pm 0.013$ & $-3.491 \pm 0.033$ & $11.820 \pm 0.020^{b}$ & [8] \\
	P167-13                    & 6.511 & $9.748  \pm 0.021$ & $12.301 \pm 0.059$ & $-2.553 \pm 0.062$ & $10.810 \pm 0.045^{b}$ & [8] \\
	P183+05                    & 6.439 & $9.854  \pm 0.019$ & $13.021 \pm 0.015$ & $-3.167 \pm 0.024$ & $11.900 \pm 0.017^{b}$ & [8] \\
	J1306+0356                 & 6.033 & $9.049  \pm 0.043$ & $12.164 \pm 0.048$ & $-3.115 \pm 0.064$ & $10.760 \pm 0.075^{b}$ & [8] \\
	J1319+0950                 & 6.135 & $9.605  \pm 0.045$ & $13.017 \pm 0.018$ & $-3.412 \pm 0.049$ & $13.245 \pm 0.227^{b}$ & [8] \\
	P231-20                    & 6.587 & $9.548  \pm 0.037$ & $13.000 \pm 0.015$ & $-3.452 \pm 0.040$ & $12.510 \pm 0.040^{b}$ & [8] \\
	P308-21                    & 6.234 & $9.528  \pm 0.024$ & $12.389 \pm 0.028$ & $-2.862 \pm 0.037$ & $10.480 \pm 0.115^{b}$ & [8] \\
	J2054-0005                 & 6.039 & $9.489  \pm 0.020$ & $12.792 \pm 0.013$ & $-3.304 \pm 0.024$ & $12.929 \pm 0.212^{b}$ & [8] \\
	J2100-1715                 & 6.039 & $9.117  \pm 0.046$ & $12.049 \pm 0.062$ & $-2.932 \pm 0.077$ & $10.060 \pm 0.130^{b}$ & [8] \\
	P323+12                    & 6.081 & $9.161  \pm 0.057$ & $11.724 \pm 0.221$ & $-2.563 \pm 0.228$ & $11.425 \pm 0.342^{b}$ & [8] \\
	J2318-3113                 & 6.587 & $9.201  \pm 0.038$ & $11.898 \pm 0.093$ & $-2.696 \pm 0.101$ & $11.610 \pm 0.035^{b}$ & [8] \\
	J2318-3029                 & 6.443 & $9.346  \pm 0.023$ & $12.799 \pm 0.010$ & $-3.452 \pm 0.025$ & $10.290 \pm 0.155^{b}$ & [8] \\
	J2348-3054                 & 6.901 & $9.248  \pm 0.044$ & $12.753 \pm 0.014$ & $-3.505 \pm 0.046$ & $12.873 \pm 0.209^{b}$ & [8] \\
	P359-06                    & 6.172 & $9.418  \pm 0.022$ & $12.204 \pm 0.041$ & $-2.786 \pm 0.046$ & $10.890 \pm 0.105^{b}$ & [8] \\
	J1120+0641                 & 7.085 & $9.083  \pm 0.039$ & $12.220 \pm 0.031$ & $-3.137 \pm 0.050$ & $12.123 \pm 0.179^{b}$ & [8] \\
	J1342+0928                 & 7.540 & $9.121  \pm 0.030$ & $11.996 \pm 0.044$ & $-2.875 \pm 0.053$ & $11.807 \pm 0.171^{b}$ & [8] \\
	\hline
	
	\multicolumn{7}{c}{\textbf{Starburst Galaxies}} \\
	\hline
	RX J094144.51+385434.8           & 1.818 & $10.362 \pm 0.130$ & $13.431 \pm 0.130$ & $-3.070 \pm 0.184$ & $13.828 \pm 0.313^{a}$ & [3] \\
	SWIRE J104704.97+592332.3 (L17)  & 1.954 & $10.301 \pm 0.130$ & $12.875 \pm 0.130$ & $-2.574 \pm 0.184$ & $13.045 \pm 0.284^{a}$ & [3] \\
	SWIRE J104738.32+591010.0 (L25)  & 1.958 & $10.079 \pm 0.130$ & $12.477 \pm 0.130$ & $-2.398 \pm 0.184$ & $12.485 \pm 0.264^{a}$ & [3] \\
	SMM J123634.51+621241.0          & 1.222 & $10.176 \pm 0.130$ & $12.602 \pm 0.130$ & $-2.426 \pm 0.184$ & $12.661 \pm 0.270^{a}$ & [3] \\
	3C 368                           & 1.130 & $9.978  \pm 0.130$ & $12.708 \pm 0.130$ & $-2.730 \pm 0.184$ & $12.809 \pm 0.275^{a}$ & [3] \\
	MIPS J142824.0+352619            & 1.325 & $10.732 \pm 0.130$ & $13.000 \pm 0.130$ & $-2.268 \pm 0.184$ & $13.221 \pm 0.290^{a}$ & [3] \\
	HDF850.1                         & 5.183 & $10.043 \pm 0.144$ & $12.813 \pm 0.067$ & $-2.770 \pm 0.159$ & $12.958 \pm 0.232^{a}$ & [9] \\
	AzTEC-3                          & 5.299 & $9.825  \pm 0.015$ & $13.041 \pm 0.087$ & $-3.216 \pm 0.088$ & $13.279 \pm 0.258^{a}$ & [10] \\
	\hline
	
	\multicolumn{7}{c}{\textbf{Composite and Other Galaxy Types}} \\
	\hline
	SDSS J100038.01+020822.4         & 1.826 & $10.041 \pm 0.130$ & $12.756 \pm 0.130$ & $-2.714 \pm 0.184$ & $12.877 \pm 0.278^{a}$ & [3] \\
	IRAS F10026+4949                 & 1.124 & $10.415 \pm 0.130$ & $13.301 \pm 0.130$ & $-2.886 \pm 0.184$ & $13.645 \pm 0.306^{a}$ & [3] \\
	SMM J22471--0206                 & 1.158 & $10.255 \pm 0.130$ & $13.301 \pm 0.130$ & $-3.046 \pm 0.184$ & $13.645 \pm 0.306^{a}$ & [3] \\
	\hline
	
	\multicolumn{7}{c}{\textbf{$z > 6$ AGN Companion Galaxies}} \\
	\hline
	P183+05C1    & 6.435 & $8.629 \pm 0.130$ & $11.910 \pm 0.075$ & $-3.281 \pm 0.150$ & $11.686 \pm 0.186^{a}$ & [8] \\
	J1319+0950C2 & 6.144 & $8.533 \pm 0.098$ & $11.577 \pm 0.115$ & $-3.045 \pm 0.151$ & $11.218 \pm 0.210^{a}$ & [8] \\
	J1342+0928C1 & 7.534 & $8.111 \pm 0.131$ & $11.228 \pm 0.144$ & $-3.117 \pm 0.195$ & $10.726 \pm 0.232^{a}$ & [8] \\
	J2318--3029C1& 6.122 & $8.955 \pm 0.100$ & $11.820 \pm 0.145$ & $-2.865 \pm 0.176$ & $11.560 \pm 0.252^{a}$ & [8] \\
	
\end{longtable}

\twocolumn[]
	
	
	\bsp	
	\label{lastpage}
\end{document}